\documentclass[oneside,openany]{aa}  
\usepackage{graphicx}
\usepackage{amsmath}
\usepackage{amssymb}
\usepackage{breqn}
\usepackage{gensymb}
\usepackage[colorinlistoftodos]{todonotes}
\usepackage[colorlinks=true, allcolors=blue]{hyperref}
\usepackage{booktabs}
\usepackage{tikz}
\usepackage{rotating}
\usepackage{natbib}
\usepackage{babel}
\usepackage{booktabs}
\usepackage{threeparttable}
\usepackage{caption}
\usepackage{subcaption}
\usepackage{adjustbox}
\usepackage{txfonts}
\usepackage{dirtytalk}
\usepackage{nicefrac, xfrac}
\usepackage{array}
\usepackage[normalem]{ulem}
\usepackage{balance}

\usepackage{xspace} 
\usepackage{soul} 

\defcitealias{W22}{W22}

\def\sgra{\object{Sgr~A*}\xspace} 
\newcommand\mdot{\dot{M}
\newcommand\mdotunit{{\rm M_{\rm \odot}/yr} }}

\newcommand\ipole{{\tt ipole}\xspace}\newcommand\bipole{{\tt bipole}\xspace}

\newcommand\dynesty{{\tt dynesty}\xspace}

\begin{document}

   \title{Fitting the light curves of Sagittarius A* with a hot-spot model}

   \subtitle{Bayesian modeling of QU loops in the millimeter band}

\titlerunning{Fitting SgrA* flares}

\authorrunning{Yfantis et al.}

   \author{A. I. Yfantis
          \inst{1}
          \and
          M. A. Mo{\'s}cibrodzka\inst{1}
          \and
          M. Wielgus\inst{2,3}
          \and
          J. T. Vos\inst{1}
          \and 
          A. Jimenez-Rosales\inst{1}
          }

    \institute{Department of Astrophysics / IMAPP, Radboud University, P.O. Box 9010, 6500 GL Nijmegen, The Netherlands\label{inst1} \\
    \email{a.yfantis@astro.ru.nl}
    \and Max-Planck-Institut für Radioastronomie, Auf dem Hügel 69, 53121 Bonn, Germany\label{2}
    \and Research Centre for Computational Physics and Data Processing, Institute of Physics, Silesian University in Opava, Bezru\v{c}ovo n\'am.~13, CZ-746\,01 Opava, Czech Republic\label{3}
    }

   \date{Received June 2023; accepted YY}

 
  \abstract
  {Sagittarius~A* (\sgra) exhibits frequent flaring activity across the electromagnetic spectrum. Signatures of an orbiting hot spot have been identified in the polarized millimeter wavelength light curves observed with ALMA in 2017 immediately after an X-ray flare. The nature of these hot spots remains uncertain.} 
   {We expanded existing theoretical hot-spot models created to describe the Sgr~A* polarized emission at millimeter wavelengths. We sampled the posterior space, identifying best-fitting parameters and characterizing uncertainties.}
 {Using the numerical radiative transfer code \ipole, we defined a semi-analytical model describing a ball of plasma orbiting Sgr~A*, threaded with a magnetic field and emitting synchrotron radiation. We then explored the posterior space in the Bayesian framework of \dynesty. We fit the static background emission separately, using a radiatively inefficient accretion flow model.}
 {We considered eight models with a varying level of complexity, distinguished by choices regarding dynamically important cooling, non-Keplerian motion, and magnetic field polarity. All models converge to realizations that fit the data, but one model without cooling, non-Keplerian motion, and magnetic field pointing toward us improved the fit significantly and also matched the observed circular polarization.}
{Our models represent observational data well and allow testing various effects in a systematic manner. From our analysis, we have inferred an inclination of $\sim155-160$~deg, which corroborates previous estimates, a preferred period of $\sim$ 90 minutes, and an orbital radius of $9 - 12.0$\, gravitational radii. Our non-Keplerian models indicate a preference for an orbital velocity of 0.6-0.9 times the Keplerian value. Last, all our models agree on a high dimensionless spin value ($a_* > 0.8$), but the impact of spin on the corresponding light curves is subdominant with respect to other parameters.}


   \maketitle

\section{Introduction}

The supermassive black hole (SMBH) at the center of our Galaxy, Sagittarius~A*, is surrounded by a hot accretion flow comprising 
relativistic plasma that emits bright synchrotron radiation. Several decades of radio as well as infrared, X-ray, and $\gamma-$ray observations of \sgra led to a detailed model for the accreting structure in the close vicinity of this SMBH  \citep[e.g.,][]{Narayan1995,Quatertaet2000,Yuan2003}. The strongly sub-Eddington luminosity of \sgra ($L_{\rm Bol} \sim 10^{-8} L_{\rm Edd}$) and high rotation measure (RM) toward the source indicate that the accretion rate is $\mdot\approx 10^{-9}-10^{-7} \mdotunit$ \citep[e.g.,][]{marrone:2006,eht:2022_paperV}. This low accretion rate for a $4\times10^6\,M_{\odot}$ SMBH is only feasible via an advection-dominated accretion flow (ADAF) mode, and the observed emission is either the ADAF itself \citep{Narayan:1997ku}, a jet launched by the ADAF \citep{Falcke:2000nai}, or both \citep{Yuan:2001dm}. Until recently, the main difficulty in elucidating the nature of the \sgra emission was the scattering screen that smears the radio view of the source up to millimeter waves (\citealt{Issaoun:2019} and references therein).

Recent observations of \sgra at millimeter wavelengths ($\lambda=1.3$\,mm, $\nu=230$\,GHz) by the Event Horizon Telescope (EHT) for the first time revealed detailed images of the event-horizon-scale emission in the Galactic center \citep{eht:2022_paperI, eht:2022_paperIII}. The EHT observations are broadly consistent with predictions of the ADAF model realized via general relativistic magenetohydrodynamic (GRMHD) simulations \citep{eht:2022_paperV}. However, when comparing a large library of GRMHD simulations (with different black hole spins and multiple ion-electron temperature ratios, magnetic field strengths, and geometries) it has been found that one of the most prominent differences between numerical models of ADAF and millimeter observations of \sgra is the variability; the models are too variable in amplitude by a factor of two compared to the light-curve data \citep{eht:2022_paperV, Wielgus2022_LC}. This discrepancy between observations and models can be associated with two main uncertainties in the models. The first uncertainty  is the unknown global geometry and stability of magnetic fields in ADAFs. The second even more problematic uncertainty is the poorly understood thermodynamics and acceleration of radiating electrons in a collisionless plasma, which is characteristic of ADAF solutions. These two main uncertainties prevent us from determining the spin of \sgra and understanding whether the observed photons originate in the accretion disk or at the jet base. Moreover, the challenges to explain all \sgra observational constraints using a single simulation are expected to increase rapidly as more detailed observational data sets are obtained. We have therefore entered an era of precision astrophysics, in which quantitative tests of theoretical predictions can be performed with event-horizon-scale astronomical observations.

Building a global model of a collisionless accretion flow and jet for \sgra using first-principle approaches, such as general relativistic particle-in-cell (GRPIC) simulations, where the electron thermodynamics and acceleration is calculated self-consistently, is currently impossible. This is mostly due to the computational complexity of the problem. GRPIC models are therefore mostly only focused on modeling outflowing magnetospheric electron-positron emission, often with unrealistic boundary conditions \citep{Crinquand2022}, or on simplified zero-angular momentum accretion models \citep{gal:2023}. Another more computationally feasible approach is to model \sgra emission using semi-analytic models \citep[e.g.,][]{Brod-Loeb2006,Broderick:2011,Younsi:2015,gravity:2018,gravity:2020a,gravity:2020b,Ball:2021,Vos:2022}. These are particularly useful because a large parameter survey can be carried out a model like this is fit to observational data.

In this work, we fit the polarized observations of \sgra near event-horizon emission reported by \citealt{W22} (hereafter \citetalias{W22}) with semi-analytic models. Our main goal is to give insight into possible conditions of collisionless plasma around \sgra, independent of the GRMHD and GRPIC interpretation. A byproduct of our fitting procedure is the determination of the spin of \sgra, its orientation on the sky, and the observer's viewing angle geometry. These can be compared to geometrical parameters inferred from different types of data such as all EHT images \citep{eht:2022_paperI,eht:2022_paperIII,eht:2022_paperV} or results from GRAVITY (e.g., \citealt{gravity:2018,gravity:2020a,gravity:2020b,gravity2023}). 

 We used polarization data from the Atacama Large Millimeter/submillimeter Array (ALMA), which observed \sgra in April 2017. We focused on 230\,GHz data that were collected immediately following an X-ray flare detected by the {\it Chandra} observatory on April 11 2017 \citep{eht:2022_paperII, Wielgus2022_LC}. The data set was presented and interpreted in \citetalias{W22}, who found that \sgra polarimetric ALMA light curves show so-called $\mathcal{Q}-\mathcal{U}$ loops, which can be well described by a model of a Gaussian hot spot threaded by a vertical magnetic field on an equatorial orbit around a Kerr black hole (\citealt{Gelles:2021}, \citealt{Vos:2022}, \citealt{Vincent2023}). In previous works, only a very limited range of the spot model parameters (mostly geometrical parameters) have been explored (\citetalias{W22}, \citealt{Vos:2022}). In this work, we couple our semi-analytic model of an orbiting spot around the Kerr black hole to a Bayesian parameter estimation framework, which allows us to carry out a large survey of the physical and geometrical parameters. The current work also presents two main extensions to our previous model. We now also simulate in an approximate way the cooling of the hot spot. Observational indications of the dynamical importance of cooling after the X-ray flare were reported in Appendix G of \citetalias{W22}. Additionally, we allow the spot to be on a non-Keplerian circular orbit, which was only briefly discussed in Appendix C of \citetalias{W22} and was studied theoretically by \citet{Vos:2022}. Our results have important implications for the GRMHD and GRPIC simulations, which are built to model \sgra, regarding the plasma conditions and dynamics.

The manuscript is structured as follows. In Section~\ref{sec:model} we briefly describe our semi-analytic model of a hot spot around \sgra. In Section~\ref{sec:fitting} we outline the data properties, our fitting framework, and the procedure. The results are presented in Section~\ref{sec:results}. We discuss the implications and provide conclusions in Section~\ref{sec:discussion}.

\begin{figure}[h!]
    \centering
    \includegraphics[width=0.921\linewidth,trim={0.1cm 0.0cm 0 0.1cm},clip]{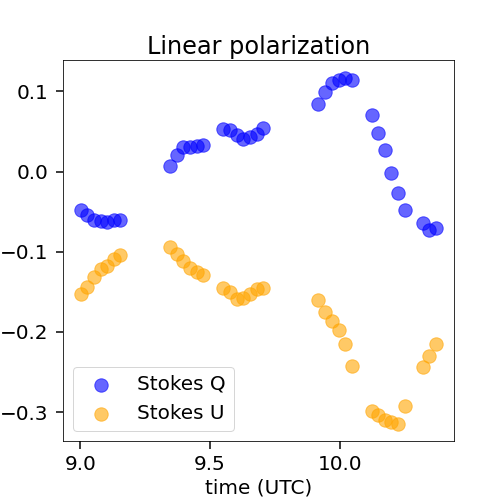}
    \caption{Observed \sgra light curves for Stokes $\mathcal Q$ and $\mathcal{U}$ from ALMA \citepalias{W22}. We only plot the points of the 35 time stamps that were used to infer the model parameters and not the 710 time stamps from the original light curve.}
    \label{fig:alma_I}
\end{figure}

\section{Model}\label{sec:model}

\begin{table*}[tbh!]
\begin{center}
     \caption{Parameters and prior ranges used for the fitting. All priors are flat, except for $n_{0,s}$, $\Theta_{0,s}$, $B_{0,s}$, $C_{\textrm{coef}}$, where we used truncated log-normal priors centered in the middle of the corresponding ranges.}
\begin{tabular}{ccl}\toprule
 \midrule 
Parameters                    & Ranges   & Description       \\ 
 \midrule 
i                    & $[90,180^\circ]$  &Viewing angle (inclination) of the observer: $i=0^\circ$ is face-on, $i=90^\circ$ is edge-on           \\
$a_*$             & $[0.0,0.9]$    &Dimensionless black hole spin        \\
$n_{0,s}$            & $[10^4,10^8]$&Electron number density at the center of the spot (cm$^{-3}$)\\
$\Theta_{0,s}$       & $[1,10^3]$     &Electron dimensionless temperature at the center of the spot \\
$B_{0,s}$             & $[0.1,10^2]$  &Magnetic field strength $(G)$\\
$r_{s}$           & $[7,15]$     &Distance of the center of the spot from black hole center ($M$)          \\
$\phi_{\rm cam}$         & $[0,360^\circ]$    &Location of the camera at the first point of observation (shifts light curves left-right)            \\
$Q_{\rm sha}$                      & $[-1,1]$    &Average background Stokes $Q$ parameter (Jy)           \\
$U_{\rm sha}$                        & $[-1,1]$      &Average background Stokes $U$ parameter (Jy)        \\
PA                             & $[-90^\circ,90^\circ]$   &Position angle of the black hole spin projected on the observer's screen, measured east of north          \\
$C_{\textrm{coef}}$                             & $[10^{-5},10^{-3}]$   &Coefficient for the exponential cooling ($M^{-1}$)           \\
$K_{\textrm{coef}}$                             & $[0.5,1.2]$   &Coefficient defining the orbital velocity with respect to the Keplerian case \\
\bottomrule 
\toprule
   \label{tab:params}
\end{tabular}
\end{center}
\end{table*}

To simulate millimeter emission from the flaring \sgra, we adoptede and further developd the semi-analytic model described in full detail by \cite{Vos:2022}. The model is inspired by work of \cite{Brod-Loeb2006} (see also \cite{Fraga-Encinas:2016nft}). It consists of a semi-analytic model of a stationary accretion flow and a time-dependent component of an orbiting bright spot around a Kerr black hole. The model is fully built within \ipole, which is a ray-tracing code for the covariant general relativistic transport of polarized light in curved space-times developed by \cite{Monika2018}. The\ipole code was used, among other applications, to generate the  EHT library of template black hole images based on GRMHD simulations (e.g., \citealt{eht:2022_paperV}). 

To calculate thermal synchrotron emission (we assumed that electrons have a relativistic thermal energy distribution function; see \citealt{Vincent2023} for a different approach), the radiative transfer model necessitates the specification of the underlying density, electron temperature, and magnetic field characteristics of the plasma. After the plasma configuration is defined, \ipole generates maps of Stokes ${\mathcal I, Q, U}$, and $\mathcal{V}$ at the selected observing frequencies. The underlying calculations employ the full relativistic radiative transfer model of \ipole, including synchrotron emission, self-absorption, and the effects of Faraday rotation and conversion. Because Sgr~A* indicates a moderate Faraday depth at 230\,GHz, with a Faraday screen most likely collocated with the compact emission region \citep{Wielgus2023}, it may be necessary to incorporate a complete radiative transfer into the model for a quantitative comparisons with observations. We self-consistently accounted for the finite velocity of light, that is, we did not employ the commonly used fast light approximation.

The model is divided into two components: a static background, modeling a radiatively inefficient ADAF disk, and an orbiting hot spot. For the background variables, we used the bg subscript, and the governing equations read
\begin{eqnarray}
    n_{e,bg}&=&n_{0,bg}\left(\frac{r}{r_g}\right)^{-1.5}\exp{\left(\frac{-\cos^2{\theta}}{2\sigma^2}\right)} \, , \nonumber\\
    \Theta_{e,bg}&=& \Theta_{0,bg} \left(\frac{r}{r_g}\right)^{-0.84}\,,\nonumber\\
    B_{bg} &=& B_{0,bg}\left(\frac{r}{r_g}\right)^{-1}\,, 
    \label{eq:bg}
\end{eqnarray}
where $n_e$ is the electron number density, $\sigma=0.3$ characterizes the disk vertical thickness, $\Theta_e=k_bT_e/m_ec^2$ is the dimensionless electron temperature, $B_{bg}$ is the magnetic field strength, $r_g = GM/c^2$ is the gravitational radius, $r$, $\theta$ are the coordinates, and $n_{0,bg}$, $\Theta_{0,bg}$, $B_{0,bg}$ are the values at the center (r=0). The field has a vertical orientation, unless specified otherwise (poloidal, or radial possibilities), with one of two possible polarities. The background ADAF is either on a fixed Keplerian orbit or deviates from it according to the parameter $K_{\textrm{coef}}$ , which corresponds to a period $P$ of 
\begin{equation}
    P=\frac{2\pi}{K_{\textrm{coef}}}\frac{r_g}{c}\left(\left(\frac{r}{r_g}\right)^{1.5} + a_*\right), 
\label{eq:P}
\end{equation} 
where $K_{\textrm{coef}}$ is a parameter controlling the Keplerianity of the orbit (for $K_{\textrm{coef}}=1$, the disk orbit is Keplerian, and for $K_{\textrm{coef}}<1$, it is sub-Keplerian).
In our model, all emission is cut below the innermost stable circular orbit (ISCO). The space-time is described with the Kerr metric, with the standard metric signature $(-\,+\,+\,+)$ and dimensionless black hole spin $a_*=Jc/GM^2$. The units are natural ($c=G=1$), hence, distance and time can be conveniently measured in units of $M$ $(M \equiv r_g \equiv r_g/c)$. We assumed a mass of Sgr A* $M=4.155 \times 10^6\,M_{\odot}$, which is between the measurements of \cite{2022A&A...657L..12G} and those of \cite{Do:2019a}. 

The second component of the model is the spot. The spot lies in a circular orbit in the equatorial plane, and its location in Kerr-Schild coordinates is defined as 

\begin{eqnarray}
\mathrm{x}_s^0 &=& t \,,\nonumber\\
\mathrm{x}_s^1 &=& r_{s}\,, \nonumber\\
\mathrm{x}_s^2 &=& \pi/2\,, \nonumber \\
\mathrm{x}_s^3 &=& 2\pi t/P \,.
\end{eqnarray}

\begin{figure*}[t]
    \centering
    \includegraphics[width=0.781\linewidth,trim={0.0cm 0.0cm 0 0.0cm},clip]
    {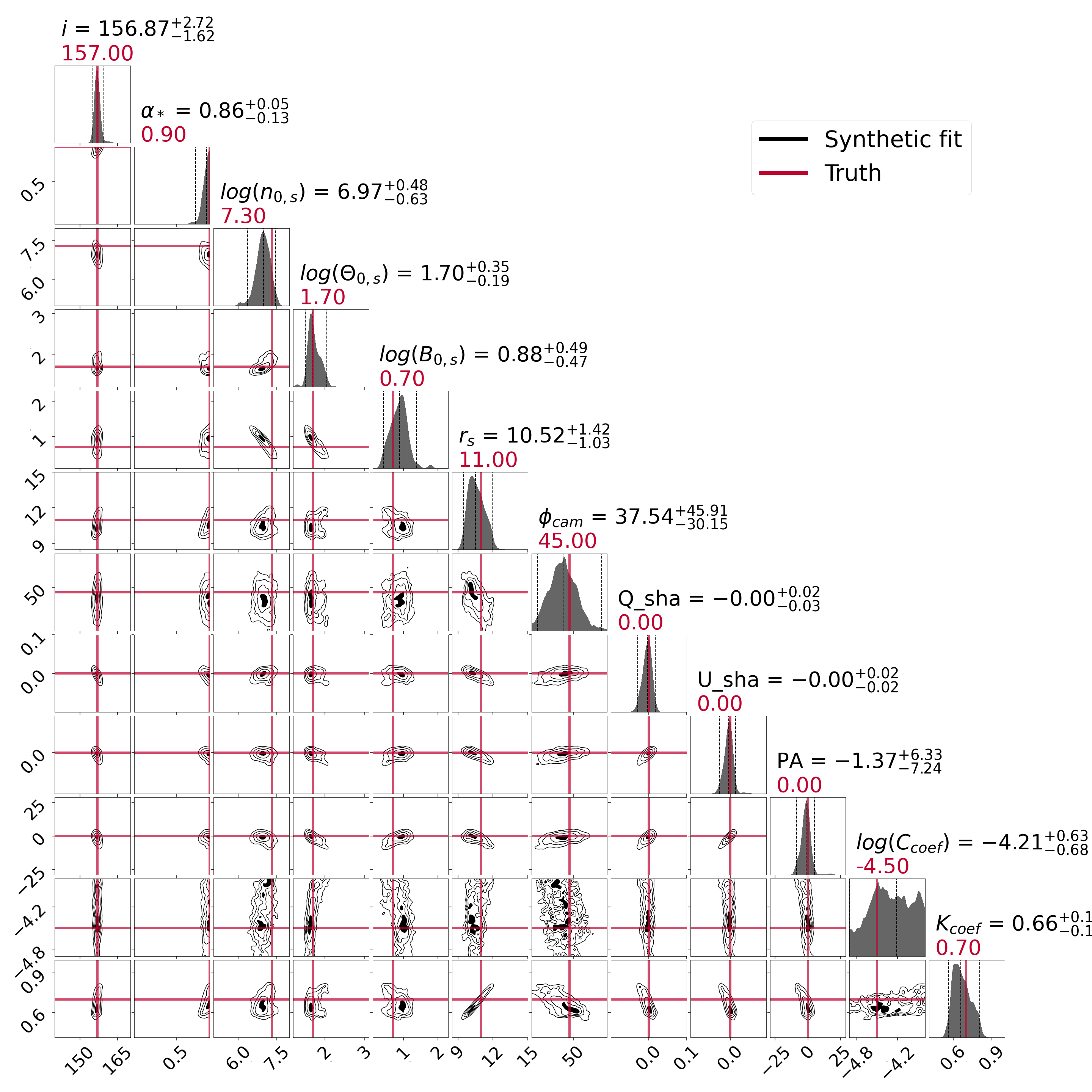}\\
    \includegraphics[width=0.29\linewidth,trim={0.0cm 42.2cm 13.5cm 0.0cm},clip]{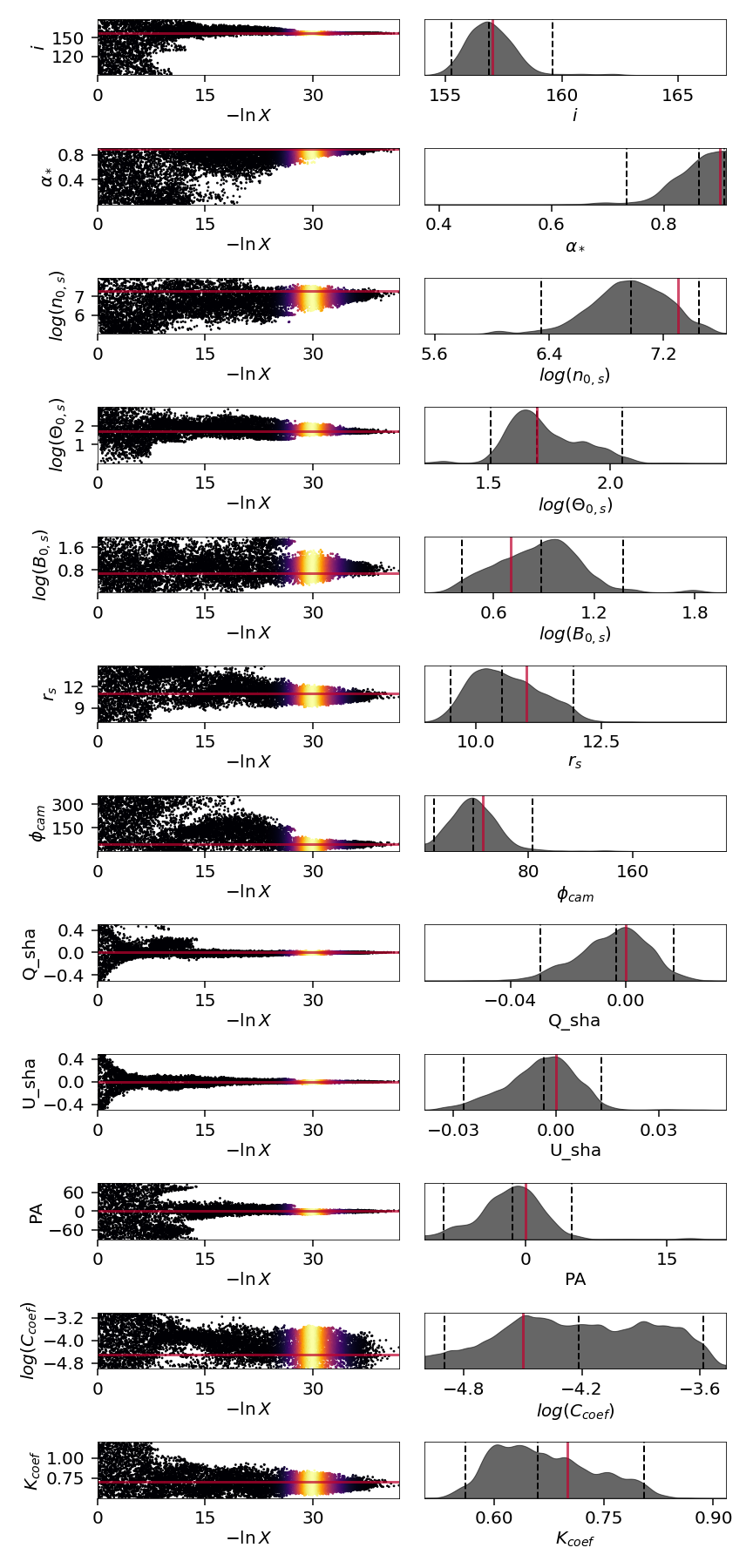}
    \includegraphics[width=0.29\linewidth,trim={0.0cm 21.1cm 13.5cm 21.1cm},clip]{figures/syn_data_plottrace_c_db_nk.png}
     \includegraphics[width=0.29\linewidth,trim={0.0cm 0.0cm 13.5cm 42.2cm},clip]{figures/syn_data_plottrace_c_db_nk.png}
    \caption{Triangle and trace plots of the synthetic light-curve fitting. Top: Triangle plot, a combination of 1D and 2D marginalized posteriors, with the dashed lines denoting the median and 2 $\sigma$ or (90$\%$) confidence levels (left and right). Similarly, the contours are for $90\%$ (outer) and then $50\%$ and $25\%$ confidence levels. The truth is marked with red lines. Bottom: Trace plots showing the evolution of particles (and their marginal posterior distributions) in 1D projections, where $\ln X$ denotes the prior volume of the parameter space. The dots are color-coded with respect to their relative posterior mass, meaning that bright yellow values are present when convergence is achieved.}
    \label{fig:syn_dat}
\end{figure*}

Here, $t$ is the coordinate time (in units of $M$), $r_{s}$ is the radial distance from the black hole center (also in units of $M$), and P is the orbital period defined in Eq. \ref{eq:P}.

The plasma density, temperature, and magnetic field strength in the spot are then described by 
\begin{eqnarray}
\begin{aligned}
& n_{e, s}=n_{0,s} \exp \left(-\frac{(\Delta x)^2}{2 R_s^2}\right), \\
& \Theta_{e, s}=\Theta_{0,s} \exp \left(-\frac{(\Delta x)^2}{2 R_s^2}\right) \exp(-C_{\textrm{coef}}(t+T_{\textrm{ini}})) ,\\
&B_{s} = B_{0,s}\exp \left(-\frac{(\Delta x)^2}{2 R_s^2}\right) \,,
\label{eq:spot}
\end{aligned}
\end{eqnarray}

where $n_{0,s}$, $\Theta_{0,s}$, and $B_{0,s}$ are the values at the center of the spot, $\Delta x$ is the Euclidean distance from the spot center to the photon emission site, connected to the distant observer's screen with a null geodesic. $R_s$ is the size of the Gaussian standard deviation, defining the size of the spot, and it remained fixed at $3M$ to limit the number of parameters. Additionally, we used a spot boundary, ignoring matter when $n_{e,s} \leq 0.01n_{0,s}$ to limit the Gaussian \say{tail}. 

We extended the spot models of \citet{Vos:2022} and allowed the Gaussian spot to change its temperature as an exponential function of time. This is a simplistic general prescription that does not aim to simulate the synchrotron cooling, but rather enabled us to explore the impact of cooling without making additional assumptions regarding its physical mechanisms. We introduce the cooling coefficient $C_{\textrm{coef}}$ and the numerical parameter $T_{\rm{ini}}=2000M$, which guarantees that the spot cooled when $C_{\textrm{coef}}>0$ (we note that \ipole tracks light backward in time from the camera position at $r=1000M$ so that the coordinate time $t$ is a negative quantity). 
The spot does not cool when $C_{\textrm{coef}}=0$.

Two numerical parameters of our model remain: the image field of view (FOV), and its resolution. We adopted FOV=40$M$$\approx 200\, \mu$as because most of the millimeter emission of \sgra \ is generated close to the black hole event horizon, with the black hole shadow ring diameter observed by the EHT corresponding to around 10-11$M$ \citep{eht:2022_paperI,eht:2022_paperVI}.
Synthetic light curves were then generated by creating snapshot images at a predefined sequence of times and integrating their intensities over the FOV into total flux densities in all four Stokes parameters. Due to large numerical cost of these calculations, we adopted a resolution of 64$\times$64 pixels per single frame.
An analysis of the effects of the image resolution on the simulated light curves can be found in Appendix~\ref{app:res}, where we justify the intermediate value of the resolution we chose. Our reasoning was that despite some fractional differences in  parts of the curves, the general behavior remains very similar and the relative differences are below the error level of the data.
\section{Two-step parameter estimation procedure}\label{sec:fitting}

ALMA observed \sgra on April 11, 2017, from 9 UTC until 13 UTC. 
Stokes $\mathcal{Q}-\mathcal{U}$ loops, interpreted as a signature of a bright spot around the black hole, were observed within the first two hours of this observing window. We fit ALMA light curves starting at 9:00 UTC until 10:30 UTC (see Fig.~\ref{fig:alma_I}). Hence, the data segment we considered begins 20 min earlier than the segment used by \citetalias{W22}. Our original data set had 710 data points, but we used 35 of these, sampling 1 out of 20 points. Within this time, the ALMA data have a few gaps in which it observed \sgra calibrators. In our modeling, we only fit Stokes $\mathcal{Q}$ and $\mathcal{U}$ light curves, but we verified Stokes $\mathcal{I}$ and ${\mathcal V}$ for consistency afterward because we assumed that during the $\mathcal{Q}-\mathcal{U}$ loop, the strongest signal of the spot is visible in the linear polarization rather than in total intensity and circular polarization, which may be dominated by the black hole shadow component originating close to the event horizon of the black hole and associated with the background image of the accretion flow \citepalias{W22}. 

To model the ALMA data, we set up a fitting framework that combined our \ipole simulations with the \dynesty software, introduced by \cite{Speagle:2019ivv} and further developed by \cite{sergey_koposov_2023_7995596}. \dynesty is a Bayesian tool capable of performing dynamic nested sampling. It was introduced for computational usage in \citet{Skilling:2006gxv}. Given our model light curves, \dynesty proved to be an excellent choice for our framework because it is compatible with low-dimensional but highly degenerate and potentially multimodal posteriors (e.g., \citealt{Palumbo:2022wnl}). 
We embedded \ipole so that the sampler called a ray-tracing operation for every log-likelihood evaluation. 

The posterior landscape of our model is very complex, and our model for the background emission (the shadow component) neither includes emission from regions within the ISCO nor has any variability. We therefore decided to simplify our fitting procedure. We introduced a two-step fitting scheme in which the spot light curve was fit separately from static background emission as follows. In our first fitting step, we fit emission from the spot alone, and in this step, the background linear polarization was modeled as the $\mathcal{Q}_{sha}$ and $\mathcal{U}_{sha}$ parameters (where $sha$ stands for shadow, following \citetalias{W22}), which were simply added to the variable light curves. In our second fitting step, we fit the static background model to the estimated pair of (${\mathcal Q}_{sha}$,${\mathcal U}_{sha}$), which represents an average emission from near the shadow of the black hole that is always visible in the millimeter waves. This split in the fitting procedure works well as long as both Faraday effects between components are negligible, for example, when the two model components are spatially separated in the images. As shown below, our spot is usually located at $r>10$$M$ with very narrow posterior distributions, so that the spot and the near horizon emission are usually well separated.

To be able to produce a fit on timescales shorter than a year, we sub-sampled the ALMA data set of \citetalias{W22} by a factor of 20, which changed cadence very little from 4\,s to 80\,s. Hence, we fit only 35 data points (or rather 70 because we took $\mathcal{Q}$ and $\mathcal{U}$ values into account separately). Because of the temporal correlation timescales involved, the subsampling has little impact on the inference power. We finally had with a scheme that performed a single log-likelihood evaluation in less than 3 seconds, using 24 cores\footnote{Using the COMA computational cluster at Radboud University.}.

The time per log-likelihood calculation requires a configuration for \dynesty that converges with the least number of iterations. To do this, we used a dynamic sampler that was introduced by \citet{Higson:2019} with 200 initial live points and 200 points per extra dynamical batch (10 in total). We used the default stopping criteria, meaning a terminating change in the log-evidence of 0.01. Additionally, we adopted a weight function that shifted the importance of the posterior exploration (rather than the evidence) to $100\%$. Last, we used \say{overlapping balls} that can generate more flexible bounding distributions. The balls were first mentioned in \citet{Buchner:2016}.  We also used a random walk as our sampling technique. This was developed by \citet{Skilling:2006gxv}. 

Our log-likelihood function is defined in a standard way as 
\begin{equation}
    \label{eq:LKLHD_CP}
    \mathcal{L}(\vec{p})=-\sum_j\frac{\left[Q_j-\hat{Q}_j(\vec{p})\right]^2}{2\sigma_{Q,j}^2} + \frac{\left[U_j-\hat{U}_j(\vec{p})\right]^2}{2\sigma_{U,j}^2}\,,
\end{equation}
where $\vec{p}$ is the vector of the parameters that are to be estimated, and $Q_j\,,U_j$ and $\hat{Q}_j(\vec{p})\,,\hat{U}_j(\vec{p})$ are the observed and modeled Stokes ${\mathcal Q}$ and ${\mathcal U}$ for every data point ($j$). The modeled quantities were directly calculated during the ray-tracing. The errors are defined as 
\begin{eqnarray}
    \sigma_{Q,j} &=& (\epsilon_{s}|Q_j| + \epsilon_{t})^{0.5}\,, \nonumber \\
    \sigma_{U,j} &=& (\epsilon_{s}|U_j| + \epsilon_{t})^{0.5}\,,
\end{eqnarray}
where $\epsilon_{s}=0.02$ is the assumed systematic error, and $\epsilon_{t}=0.01$ represents thermal noise. For the characterization of uncertainties in the ALMA data see \citealt{Wielgus2022_LC}, where the formal ALMA light-curve uncertainties were found to be underestimated. The reduced $\chi^2$ then naturally follows as

\begin{equation}
\chi_{\rm eff}^2 = \frac{\chi^2}{n_{d}-n_{f}}
    \label{eq:x_sq},
\end{equation}
where $\chi^2 = -2\mathcal{L}$, $n_d = 70 \,, (2\times35)$ is the number of data points, and $n_f = 10,\,11$, or $12$ is the number of degrees of freedom, which is equal to the number of parameters that were fit simultaneously.

A comprehensive list of all the parameters and their ranges used in our fitting pipeline, referred to as \texttt{bipole} in the next sections, is shown in Table~\ref{tab:params}. For $n_{0,s},\,\Theta_{0,s}$, and $B_{0,s}$ we sampled the log space with Gaussian priors centered in the middle of their log ranges (i.e., log-normal priors).  For the remaining parameters we considered, we assume flat priors.


Finally, before fitting the real data, we performed a controlled \texttt{bipole} run using synthetic data to demonstrate the ability of our pipeline to recover the truth value. The total number of fitted parameters was the same as for the real data (see Section~\ref{sec:results}) with the same priors.
The results are presented in Fig.~\ref{fig:syn_dat}. On the left side, we present the trace plot, showing the evolution of the live points throughout the run. On the right side, we show the corner (or triangle) plot, showing the single and joint distributions for all parameters.  All truth values, indicated with red lines, are comfortably covered by the posteriors.
Additionally, we observe a correlation between the  plasma parameters ($n_{0,s}\, \Theta_{0,s},\, B_{0,s}$) that is expected for synchrotron emission, a correlation between $r_s$ and $K_{\rm coef}$, which is expected given the period relation, and an overall less strictly constrained $C_{\rm coef}$ posterior, which is expected given the volatility of this parameter for small changes in plasma parameters.

\section{Results}\label{sec:results}

\subsection{Spot fitting}
\let\cleardoublepage\clearpage

\begin{table}[tbh!]
\begin{center}
     \caption{Models used to fit Sgr~A* ALMA data. The default magnetic (db) refers to the field polarity aligned with the spin axis of the black hole, pointing away from the observer for $i>90^\circ$.}
\begin{tabular}{llccc}\toprule
 \midrule 
Model ID &$B_{\rm field}$ &Cooling &Keplerian &Parameters\\ 
\midrule
nc\_db_k  & default     & no      & yes  &10\\  
nc\_fb\_k_k & flipped     & no      & yes  &10\\
c\_db_k   & default     & yes     & yes  &11\\
c\_fb_k  & flipped     & yes     & yes  &11\\
nc\_db_nk  & default     & no      & no  &11\\
nc\_fb\_nk & flipped     & no      & no  &11\\
c\_db_nk   & default     & yes     & no  &12\\
c\_fb_nk  & flipped     & yes     & no  &12\\
\bottomrule 
\toprule
\label{tab:models}
\end{tabular}
\end{center}
\end{table}

We ran eight models that we list in Table~\ref{tab:models} to fit the spot component to the ALMA observational data. The set of models was composed of all combinations of three criteria: 1) no cooling or fitting the cooling parameter $C_{\rm coef}$; 2) prescribed default or flipped magnetic field orientation; in the default models, the magnetic field vector was aligned with the black hole spin, which is also aligned with the angular momentum vector of the hot spot, which points away from us for the prior space of inclination ($i>90^\circ$); and 3) assumed Keplerian motion, or fitting  a non-Keplerianity parameter $K_{\rm coef}$. The models were labeled to contain the information of the specific case. The first part refers to cooling (nc being no cooling), the second part to $B_{\textrm{field}}$, and the last part to a Keplerian orbit (nk being the non-Keplerian orbit). We report below the posteriors from each run, discuss the goodness of fit, and show the produced light curves with the best parameters from every model.


\begin{figure*}[h!]
    \centering
    \includegraphics[width=0.781\linewidth,trim={0.2cm 0.0cm 0 0.2cm},clip]{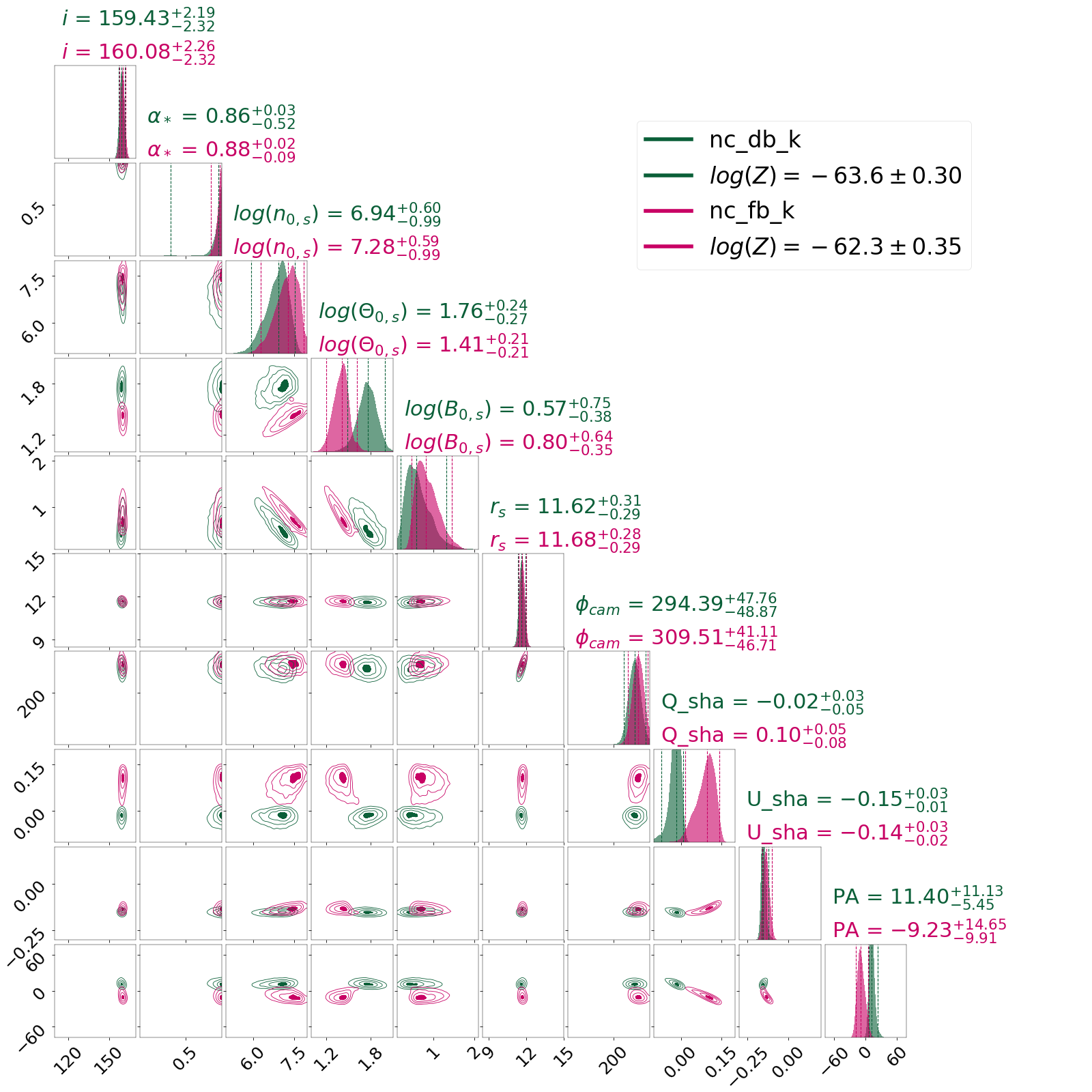} \\
    \includegraphics[width=0.441\linewidth,trim={0.1cm 0.0cm 0 0.1cm},clip]{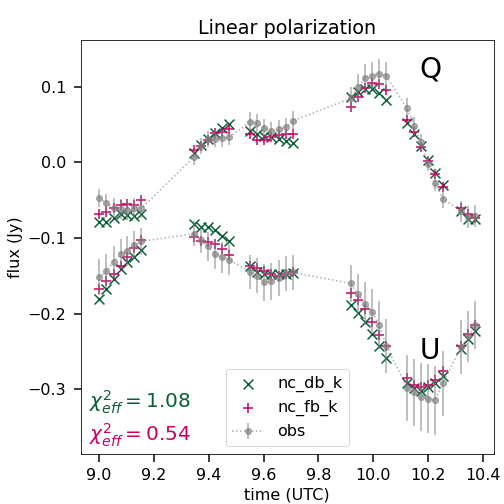} 
    \includegraphics[width=0.441\linewidth,trim={0.1cm 0.0cm 0 0.1cm},clip]{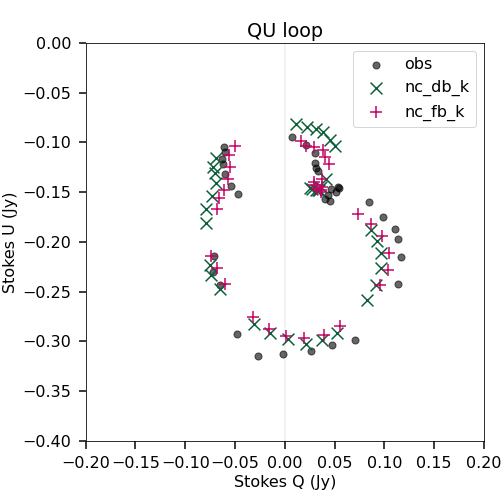}
    \caption{Triangle plot and light curves for the Keplerian spot models without cooling ($K_{\rm coef}=1$, $C_{\rm coef}=0$). Top: Triangle plot of the posterior distributions for all parameters. The results for the two magnetic field polarities are shown in different colors, and their respected log(Evidence) values are reported in the label.
    Bottom: Best-fit model $\mathcal{Q}-\mathcal{U}$ light curves and $\mathcal{Q}-\mathcal{U}$ loops (color points) overplotted with ALMA observational data (gray points). The $\chi_{\rm eff}^2$ of each light curve is visible in the bottom left corner.}
    \label{fig:corner_mA}
\end{figure*}

\begin{figure*}[h!]
    \centering
    \includegraphics[width=0.791\linewidth,trim={0.0cm 0.0cm 0 0.0cm},clip]{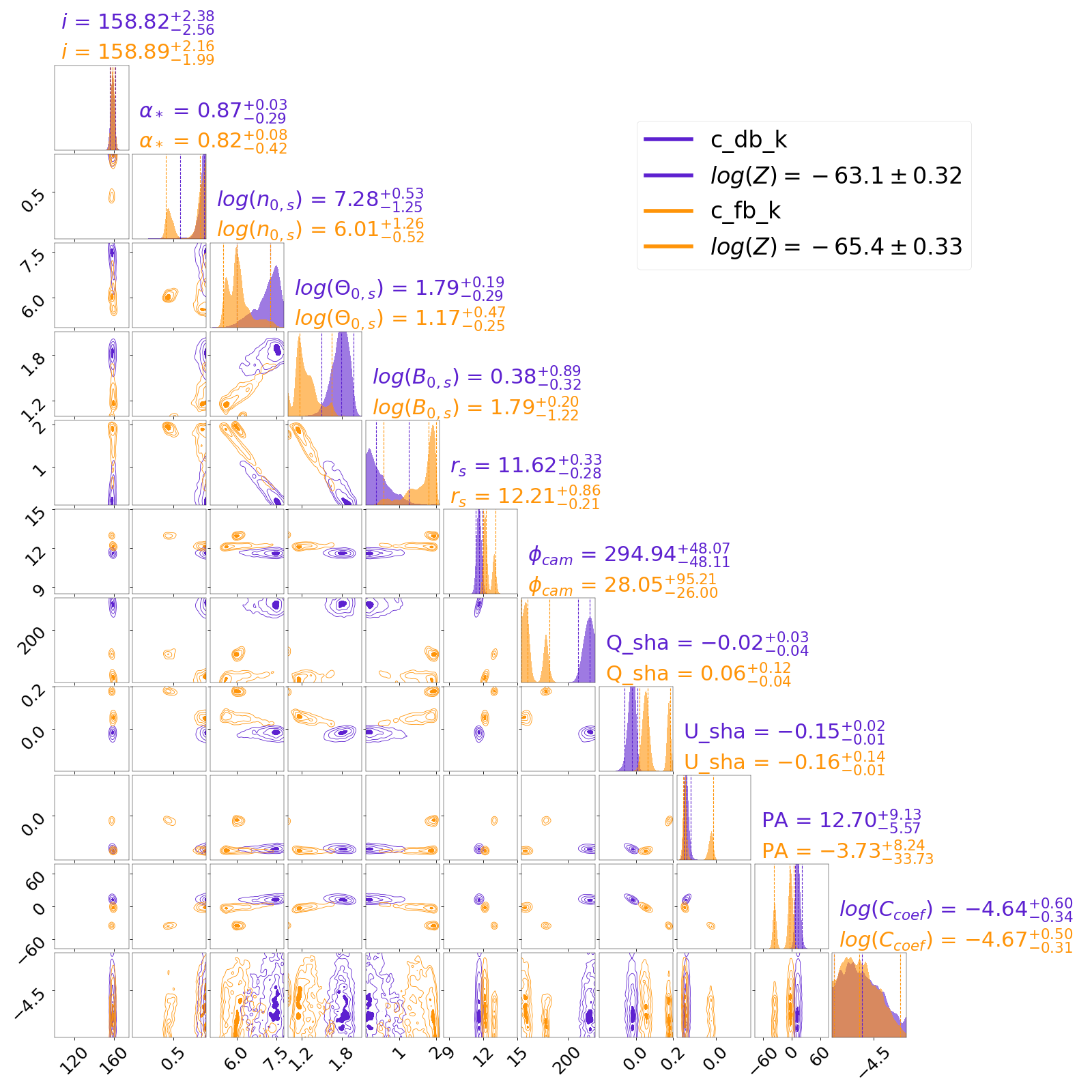}\\
    \includegraphics[width=0.451\linewidth,trim={0.1cm 0.0cm 0 0.1cm},clip]{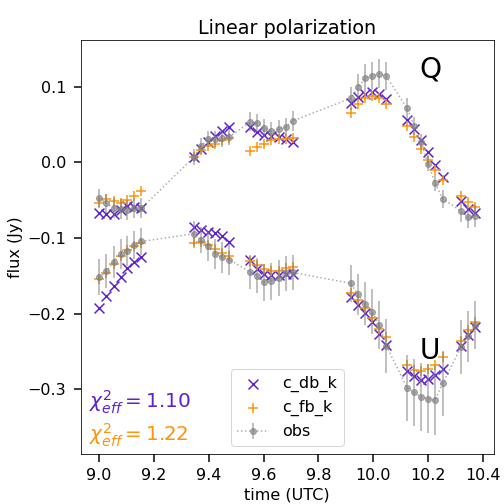}
    \includegraphics[width=0.451\linewidth,trim={0.1cm 0.0cm 0 0.1cm},clip]{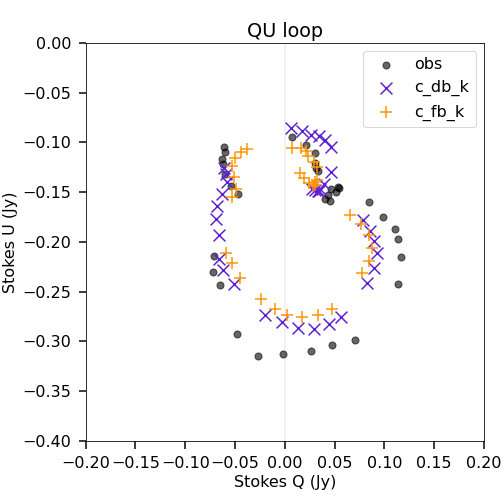}
    \caption{ Same as in Fig.~\ref{fig:corner_mA}, but for Keplerian models with cooling ($K_{\rm coef}=1$, $C_{\rm coef}\neq 0$).}    
    \label{fig:corner_cool}
\end{figure*}

\begin{figure*}[h!]
    \centering
    \includegraphics[width=0.791\linewidth,trim={0.0cm 0.0cm 0 0.0cm},clip]{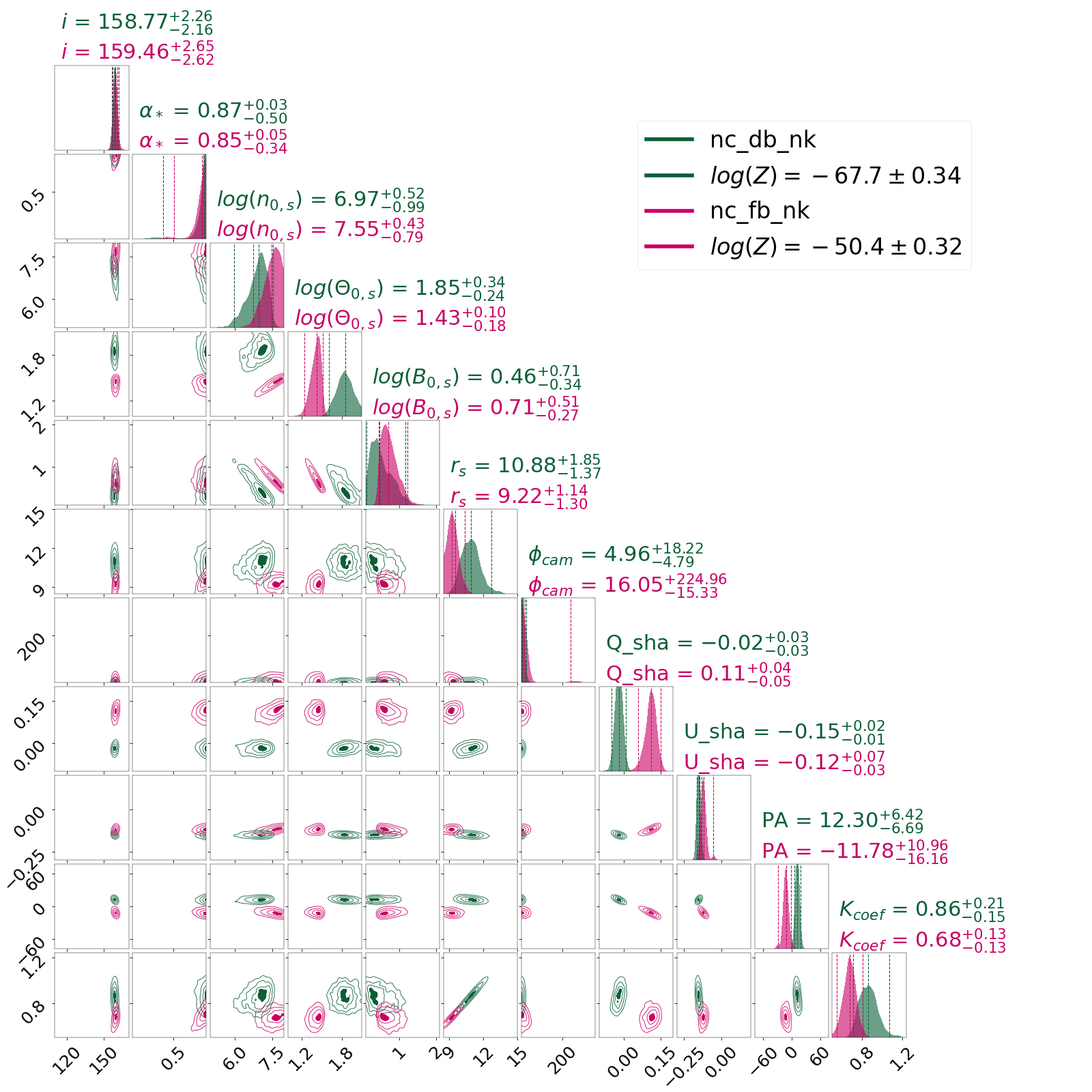}
    \includegraphics[width=0.451\linewidth,trim={0.1cm 0.0cm 0 0.1cm},clip]{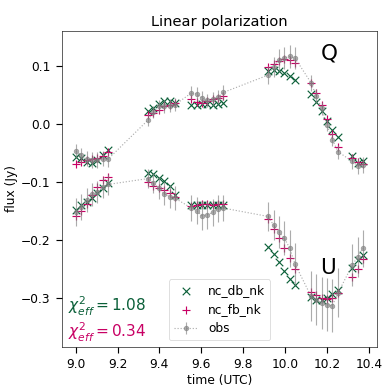}
    \includegraphics[width=0.451\linewidth,trim={0.1cm 0.0cm 0 0.1cm},clip]{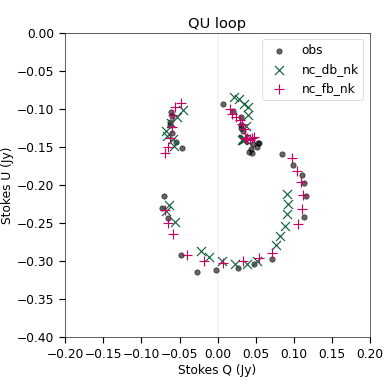}\\
    \caption{Same as in Fig.~\ref{fig:corner_mA}, but for non-Keplerian models without cooling ($K_{\rm coef} \neq 1$,$C_{\rm coef} = 0$).}    
    \label{fig:corner_nk}
\end{figure*}

\begin{figure*}[h!]
    \centering
    \includegraphics[width=0.771\linewidth,trim={0.0cm 0.0cm 0 0.0cm},clip]{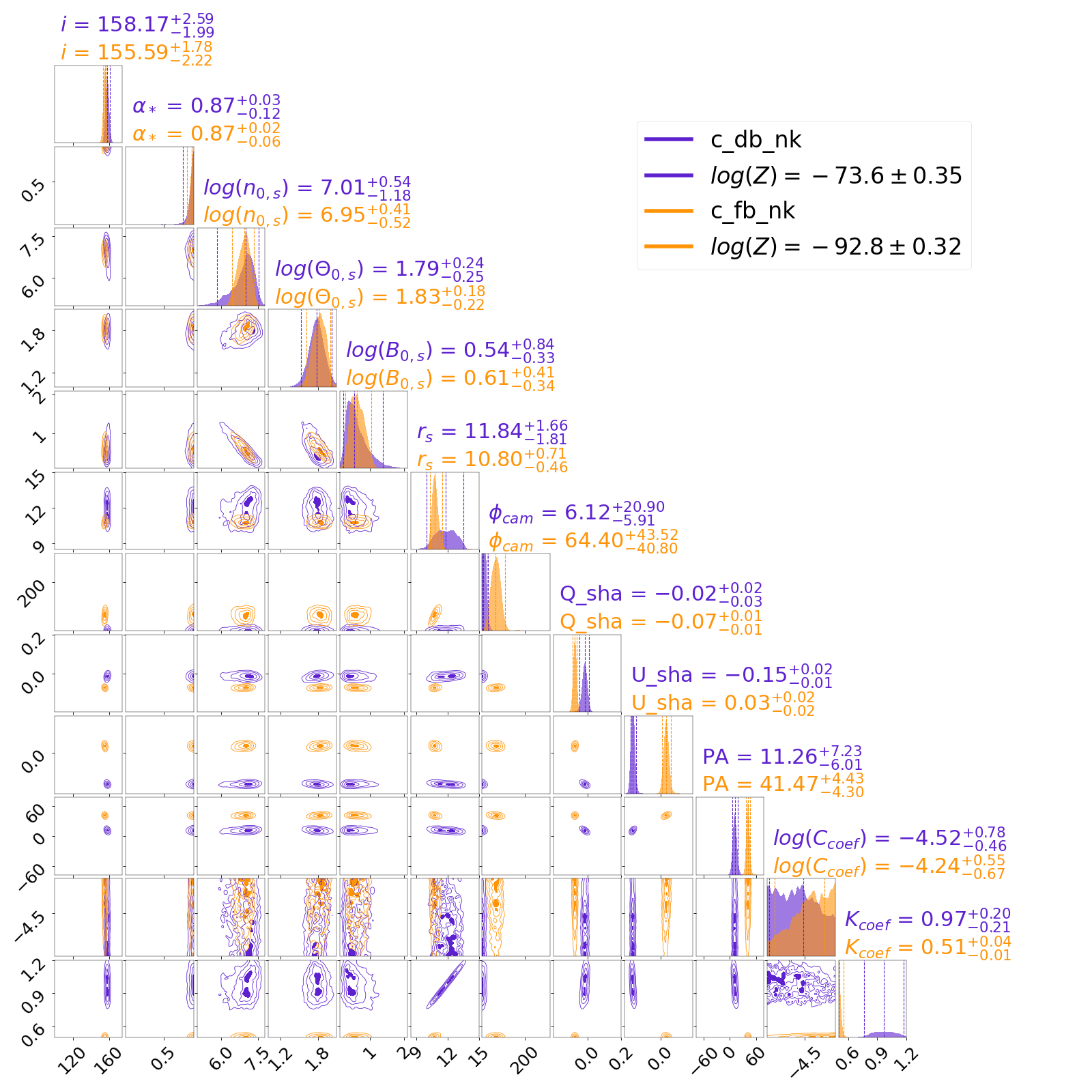}\\
    \includegraphics[width=0.451\linewidth,trim={0.1cm 0.0cm 0 0.1cm},clip]{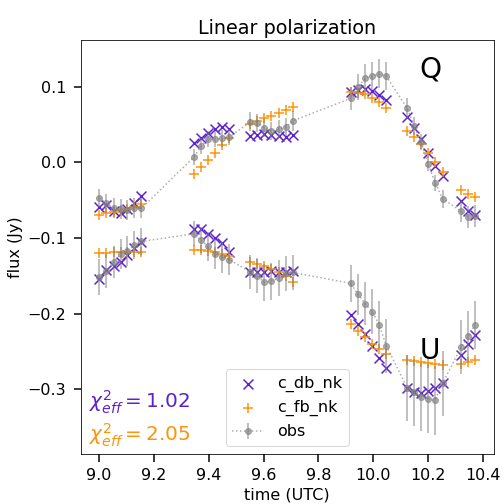}
    \includegraphics[width=0.451\linewidth,trim={0.1cm 0.0cm 0 0.1cm},clip]{figures/lc_QUlc_c_db_k.png}
    \caption{Same as in Fig.~\ref{fig:corner_mA}, but for non-Keplerian models with cooling ($K_{\rm coef} \neq 1$, $C_{\rm coef} \neq 0$).}    
    \label{fig:corner_c_nk}
\end{figure*}

In Fig.~\ref{fig:corner_mA} (top part) we show the posterior distributions for all parameters for two different magnetic field orientations in Keplerian models without cooling (nc\_db\_k, nc\_fb\_k). The first important feature to note is the agreement in the value of inclination for both models, which matches the fiducial model of \citetalias{W22}, who assumed an inclination of 158 deg. Second, the values of $r_s$ are identical in the models ($r_s=11.65M$ and with narrow posteriors), but larger than those of \citetalias{W22}, who assumed $10-11M$. This has a direct effect on the loop period (Eq. \ref{eq:P}, for $r_s \sim 10$$M$ and fixed Keplerian motion spin has little effect), producing a period of $T=88$min, which is longer than the 70\,min proposed in \citetalias{W22}. This is further discussed in the next section \ref{sec:discussion}. The black hole spin is constrained at the upper end of the prior space ($a_* \sim 0.9$); this is further discussed in Section~\ref{sec:discussion}. The plasma density, electron temperature, and magnetic field strengths are normally distributed around $n_{0,s}\sim10^7\,{\rm particles}/$cm$^{-3}$, and $\Theta_{0,s}\sim 40$, $B_{0,s}\sim 5$G. There is also a correlation between $n_{0,s}$, $\Theta_{0,s}$ and $B_{0,s}$ similar to the synthetic data fit.  
The PA values lie around zero. 
Between the default and flipped $B_{\rm field}$ fits, some parameters exhibit noticeable shifts, among which $Q_{\rm sha}$ is most significantly affected (see Section~\ref{subsec:bg}). From the reported log(Evidence) alone, there is no strong preference for either of the two $B_{\rm field}$ orientations.
In the bottom part of Fig.~\ref{fig:corner_mA}, we present the light curves and the $\mathcal{Q}-\mathcal{U}$ loop using the model with the highest likelihood from each run as parameters for the ray-tracing. In the lower left corner of the light curve, we report the lowest $\chi_{\rm eff}^2$ value. The nc\_fb\_k model has  $\chi^2_{\rm eff}=0.54$, that is,  half of the $\chi^2_{\rm eff}=1.08$ for nc\_db\_k model. 

For the second family of models used in the fitting pipeline (c\_db\_k, c\_fb\_k), we incorporated an additional parameter, namely $C_{\textrm{coef}}\neq0$, which permits the time evolution of the electron temperature values in the spot according to Eq.~\ref{eq:spot}. 
In Fig.~\ref{fig:corner_cool} (top panels) we show the posterior distributions for all free parameters of this model. The inclination is consistent between the two orientations of $B_{\rm field}$ and with the no-cooling models nc\_db\_k and nc\_fb\_k. In the case of c\_db\_k, the remaining parameters are also very similar to the models without cooling, but the plasma parameters show longer tails. This effect arises because $C_{\textrm{coef}}$ spans an order of magnitude with $2 \sigma$ confidence. In contrast, in the case of c\_fb\_k, the remaining distributions exhibit many differences. First, there is a prominent bimodal character to the posteriors that affects all parameters except for the plasma parameters. Second, the plasma parameters do not remain similar either, with a modification of the synchrotron emission caused by higher $B_{\textrm{field}}$ values $(\sim60G)$ and lower densities and temperatures ($10^6$ and $10$, respectively). Of the two realizations of c\_fb\_k, the realization with high spin seems more likely because it produces a period of $T=91$min, while the other realization produces $T=101$\,min. The cooling coefficient is the same as the default $B_{\textrm{field}}$ ($C_{\textrm{coef}}\sim 10^{-5}-10^{-4}$), producing a drop in temperature ranging from $0.01\%-2\%$ during the simulation (light-curve duration of $\sim$1.5h). 
In the bottom panels of Fig.~\ref{fig:corner_cool}, we present the model c\_b and c\_fb light curves and the $\mathcal{Q}-\mathcal{U}$ loop using the highest likelihood models from each run as parameters for the ray-tracing. 
Their $\chi_{\rm eff}^2$ is reported in the lower left corner of the figure. The two models with cooling again prefer high black hole spins, while $\chi_{\rm eff}^2 \sim 1.1$ and $\chi_{\rm eff}^2 \sim 1.22$ here are twice higher than the values reported for model nc\_fb\_k. Models without and with cooling have similar Evidence.

In Fig.~\ref{fig:corner_nk} (top panels) we show the posterior distributions for all free parameters for the non-Keplerian model without cooling (models nc\_db\_nk and nc\_fb\_nk with $K_{\textrm{coef}}\neq1$, $C_{\textrm{coef}}=0$). The differences between the two $B_{\textrm{field}}$ orientations are enhanced. Model nc\_fb\_nk produces a strikingly high log(Evidence)=$-50$ that distinguishes it among all other models, with log(Z)$< -63$. The consistency in inclination and spin values persists, but the values for the spot radius are significantly lower ($r_s=9-10M$). This is a direct effect of non-Keplerian motion expressed as subunit values of $K_{\textrm{coef}}=0.6-1.0$.
The light curves are shown in the bottom panels of Fig.~\ref{fig:corner_nk} together with the $\chi_{\rm eff}^2$ values of each model. The low value of $\chi_{\rm eff}^2=0.34$ together with the log(Evidence) and the smooth posteriors make nc\_fb\_nk our best-bet model.

In Fig.~\ref{fig:corner_c_nk} (top panels) we show the posterior distributions for all free parameters for the non-Keplerian model with cooling (models c\_b\_nk and c\_fb\_nk  with $K_{\textrm{coef}}\neq1 $, $C_{\textrm{coef}}\neq0$,). The inclination and spin are still the same, but the two $B_{\textrm{field}}$ orientations again behave very differently when cooling is included. While both cases have a similar $C_{\textrm{coef}}$ range, $K_{\textrm{coef}}$ behaves much differently than in the nc\_db\_nk and nc\_fb\_nk cases. The c\_db\_nk has a wide $K_{\textrm{coef}}=0.8-1.2$ that correlates well with $r_s$, while c\_fb\_nk has a narrow $K_{\textrm{coef}}=0.5$ that in combination with $r_s=10.8 M$ does not reproduce the required period. The log(Evidence) for both models is lower than in any of the remaining models, but in particular, for c\_fb\_nk the Evidence together with the $\chi_{\rm eff}^2=2.05$ make the fit most unreliable. The light curves are shown in the bottom panels of Fig.~\ref{fig:corner_c_nk}.

The orbital periods of most of the best-fit models (with just one exception) fall into a narrow range between 87 and 93 minutes, which corresponds to the duration of our fit light curve (1.5\,h). For comparison, the model c\_fb\_nk has a much longer period of approximately 155 minutes. All orbital periods of the spots are reported in Table~\ref{tab:logz} in Section~\ref{sec:discussion}.

\subsection{Cross-checking with other Stokes parameters}
\label{subsec:cross-check}

In this section, we report the behavior of the Stokes $\mathcal{I}$ and $\mathcal{V}$ light curves for the best-fit realizations of the eight discussed models. The goal is to verify the consistency of the models with the remaining observational information that was not used in the fitting procedure because of the model limitations listed in Sect.~\ref{sec:fitting}. 

\begin{figure*}[h!]
    \centering
    \includegraphics[width=0.231\linewidth,trim={0.1cm 0.0cm 0 0.1cm},clip]{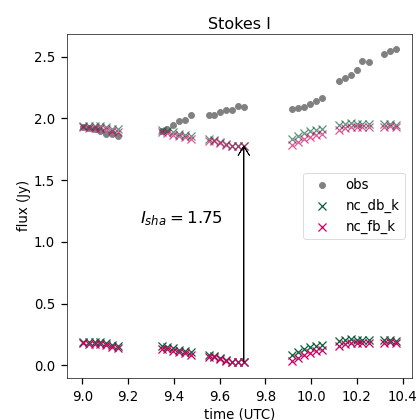}
    \includegraphics[width=0.231\linewidth,trim={0.1cm 0.0cm 0 0.1cm},clip]{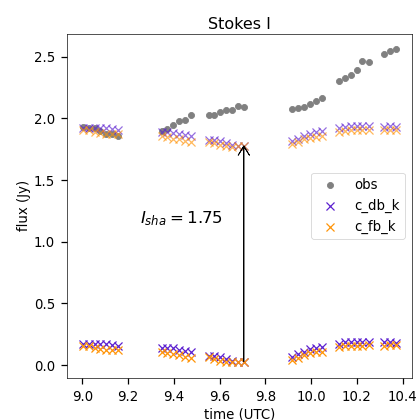} 
    \includegraphics[width=0.231\linewidth,trim={0.1cm 0.0cm 0 0.1cm},clip]{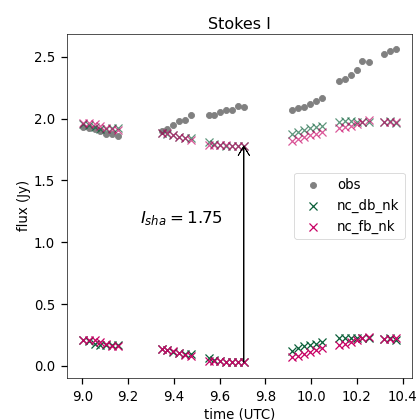}
    \includegraphics[width=0.231\linewidth,trim={0.1cm 0.0cm 0 0.1cm},clip]{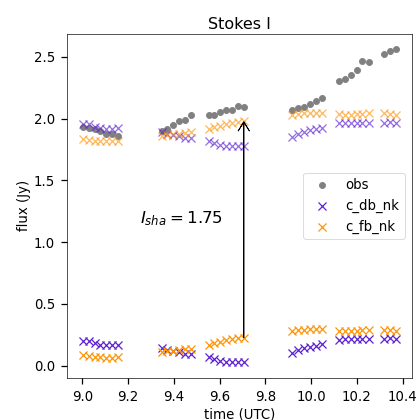}\\
    \includegraphics[width=0.231\linewidth,trim={0.1cm 0.0cm 0 0.1cm},clip]{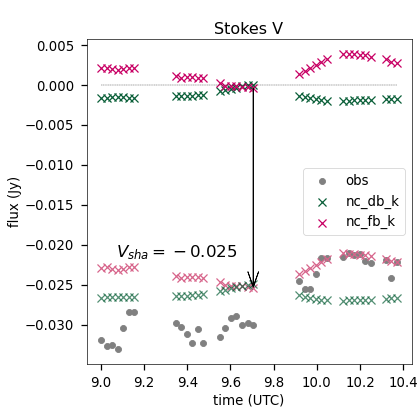}
    \includegraphics[width=0.231\linewidth,trim={0.1cm 0.0cm 0 0.1cm},clip]{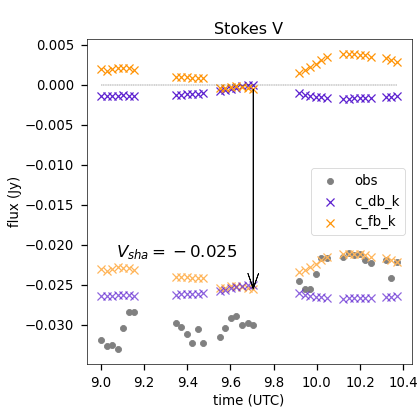}
    \includegraphics[width=0.231\linewidth,trim={0.1cm 0.0cm 0 0.1cm},clip]{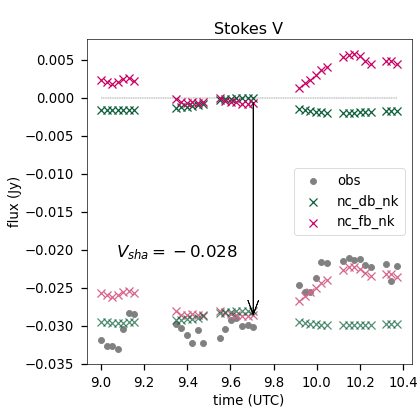}
    \includegraphics[width=0.231\linewidth,trim={0.1cm 0.0cm 0 0.1cm},clip]{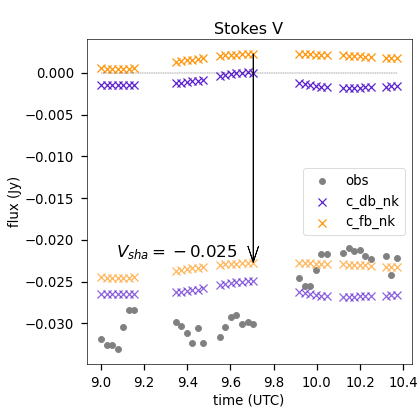}
    \caption{Stokes $\mathcal{I},\, \mathcal{V}$ light curves for all models compared to the observational data from \citet{Wielgus2022_LC} and \citetalias{W22} (obs). The arrows represent a constant background component to facilitate comparison with the real data.}  
    \label{fig:lc_IV}
\end{figure*}

In the top row of Fig.~\ref{fig:lc_IV}, we present the Stokes $\mathcal{I}$ light curves for all spot models. The shift (sha) of the light curves is just for visual aid, and it accommodates the contribution we expect from the background (the accretion disk). We did not fit for the best \say{shadow} (sha) values to Stokes $\mathcal{I}$ (and $\mathcal{V}$, see below). We observe only a small variation in Stokes $\mathcal{I}$ in all models. This is expected since Stokes $\mathcal{I}$ should be largely dominated by the background emission, and any variations could be due to the background variability, which we did not model here. 

In the bottom row of Fig.~\ref{fig:lc_IV}, we show the light curves for Stokes $\mathcal{V}$. All models with flipped $B_{\textrm{field}}$ except for one model exhibit a behavior that resembles the data more closely than the default $B_{\textrm{field}}$ models. When we apply a shift with the same shadow concept as for Stokes $\mathcal{I}$, the coincidence with the data is evident. In particular, the best-bet model nc\_fb\_nk recovers the observed Stokes $\mathcal{V}$ best, which is a strong independent argument in support of the model. 

\subsection{Background emission}\label{subsec:bg}

Here, we report the second part of our two-step fitting algorithm. We fit the $\mathcal{Q}_{\textrm{sha}}$, $\mathcal{U}_{\textrm{sha}}$ values from every model (reported in Table~\ref{tab:QU_sha}) with the ADAF model described by Eqs.~\ref{eq:bg}. Our goal was to obtain the density, temperature, and magnetic field strengths of the background component. To fit the plasma parameters of the background emission, we fixed all the geometric parameters ($a_*$, $i$, PA), plus $K_{\rm coef}$ for non-Keplerian models, at the values obtained in the spot-fitting (during the first step). The choice to simulate the disk with the same $K_{\textrm{coef}}$ as the spot is an assumption because other possibilities could be meaningful as well, for instance, a Keplerian disk with a sub-Keplerian spot. The exclusion of the radiation originating inside the ISCO constitutes an additional limitation.  Last, we note that since the background is static, we only need one point (or rather two, for $\mathcal{Q}$ and $\mathcal{U}$) to make the fit.

\begin{table}
\begin{center}
     \caption{ $\mathcal{Q}_{\textrm{sha}}$, $\mathcal{U}_{\textrm{sha}}$ values for all models obtained in the first fitting step, as well as the average values of $\mathcal{Q}$ and $\mathcal{U}$ from the ALMA data.}
\begin{tabular}{lcc}\toprule
 \midrule 
Model ID &$Q_{\textrm{sha}}$ (Jy) & $U_{\textrm{sha}}$ (Jy) \\ 
\midrule
nc\_db\_k  & $-0.02$     & $-0.15$    \\  
nc\_fb\_k & $0.1$      & $-0.14$    \\
c\_b   & $-0.02$      & $-0.15$     \\
c\_fb  & $0.06$      & $-0.16$     \\
nc\_db\_nk  & $-0.02$  & $-0.15$     \\
nc\_fb\_nk & $0.11$  & $-0.12$      \\
c_b_nk   & $-0.02$   & $-0.15$     \\
c\_fb_nk  & $-0.07$  & $0.03$     \\
\midrule
ALMA average & $0.015$ & $-0.18$ \\        
\bottomrule 
\toprule
\label{tab:QU_sha}
\end{tabular}
\end{center}
\end{table}

\begin{figure*}[h!]
    \centering
    \includegraphics[width=0.48\linewidth,trim={0.1cm 0.0cm 0 0.1cm},clip]{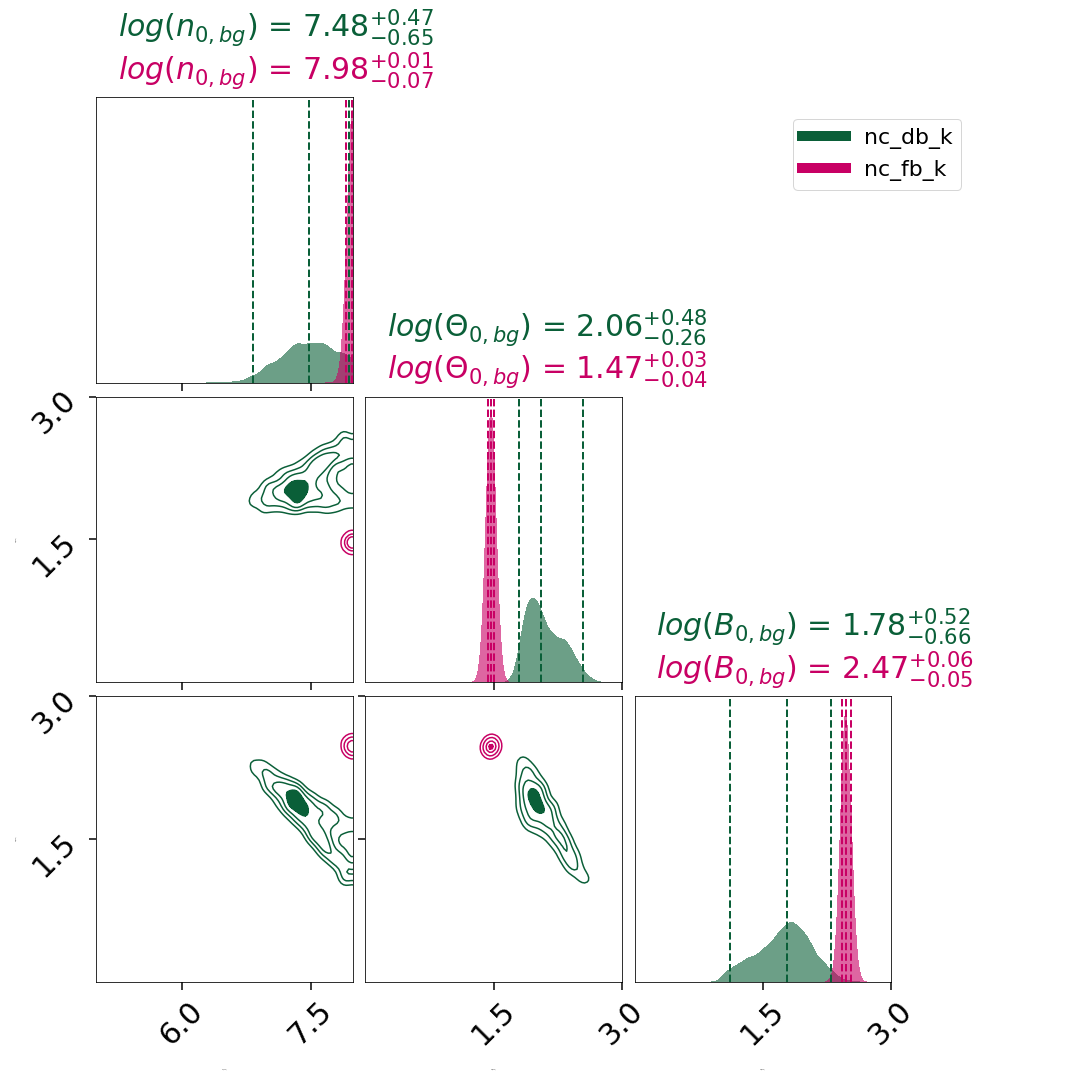} 
    \includegraphics[width=0.48\linewidth,trim={0.1cm 0.0cm 0 0.1cm},clip]{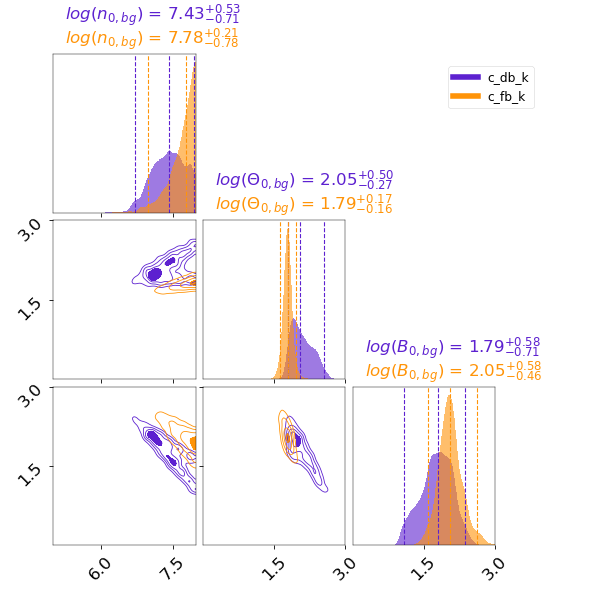}\\
    \includegraphics[width=0.48\linewidth,trim={0.1cm 0.0cm 0 0.1cm},clip]{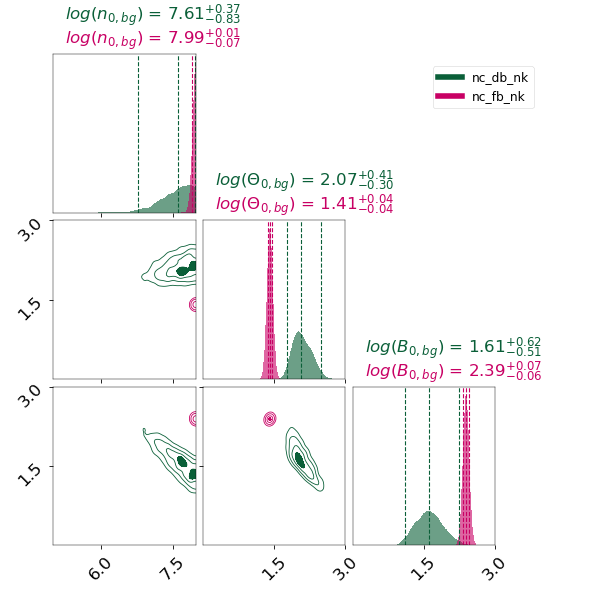} 
    \includegraphics[width=0.48\linewidth,trim={0.1cm 0.0cm 0 0.1cm},clip]{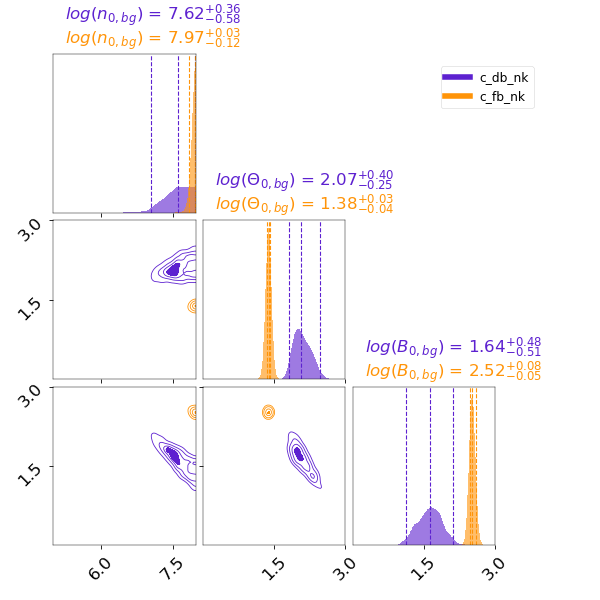}\\
    \caption{ Triangle plots for the background solution. The geometric parameters ($i,a$, PA), and $K_{\rm coef}$ for nk cases were fixed at the values of the initial runs for the two $B_{\textrm{field}}$ orientations (color-coded) and without (left panels) and with cooling  (right panels).}   
    \label{fig:corner_bg3}
\end{figure*}

In Figure~\ref{fig:corner_bg3} we show the posteriors from the aforementioned background models. All runs converge, but there is a clear difference between the two $B_{\rm field}$ orientations, which is the level of correlation. The default $B_{\rm field}$ has higher correlations, which is an indication that it might describe the background emission better. The values for all three parameters are roughly higher by an order of magnitude than what we expect, but this is most likely due to the lack of radiation from within the ISCO.

Using the peaks of the posteriors from the three parameter runs, we calculated Stokes $\mathcal{I}_{\textrm{BG}}$, $\mathcal{V}_{\textrm{BG}}$ produced from the background disk. Our motivation was to test whether any of the Stokes values match the reported $\mathcal{I}_{\textrm{sha}}$, $\mathcal{V}_{\textrm{sha}}$ values used in Fig. \ref{fig:lc_IV}. The values are reported in Table \ref{tab:IV}. All values from the default $B_{\rm field}$ models match the shadow components fairly well. Moreover, the models with a flipped magnetic field orientation produce opposite values for Stokes $\mathcal{V}_{\textrm{BG}}$. We comment on the implications of these results in Section \ref{sec:discussion}. 
Overall, this second step does not offer much inferring power at this point, but it serves as a pointer for the preferred $B_{\rm field}$ orientation and is a sanity check and validation of our overall approach of splitting the fitting into two steps.

\begin{table}
\begin{center}
     \caption{ $\mathcal{I}_{\textrm{BG}}$, $\mathcal{V}_{\textrm{BG}}$ values for all models obtained in the second fitting step (three parameters) are directly compared to the $\mathcal{I}_{\textrm{sha}}$, $\mathcal{V}_{\textrm{sha}}$ used for visual aid in Fig. \ref{fig:lc_IV}, and to the average values of $\mathcal{I}$ and $\mathcal{V}$ from the ALMA data. The average of $\mathcal{I}$ is from the first 10 points since the later increase in Stokes $I_{\textrm{sha}}$ is argued to occur due to the time evolution of the background or the spot-background interactions.}
     
\begin{tabular}{lcccc}\toprule
 \midrule 
Model ID &$I_{\textrm{BG}}$ (Jy) & $V_{\textrm{BG}}$ (Jy) &$I_{\textrm{sha}}$ (Jy)& $V_{\textrm{sha}}$ (Jy) \\ 
\midrule
nc\_db\_k  & $2.01$     & $-0.036$ & $1.75$ & $ -0.025$  \\  
nc\_fb\_k & $4.27$      & $+0.34$ &$1.75$ &$-0.025$   \\
c\_db\_k   & $1.73$      & $-0.028$ & $1.75$ & $ -0.025$   \\
c\_fb\_k  & $3.63$      & $+0.025$  & $1.75$ & $-0.025$  \\
nc\_db\_nk  & $1.24$  & $-0.022$ & $1.75$ & $-0.028$   \\
nc\_fb\_nk & $2.86$  & $+0.24$ & $1.75$ & $-0.028$     \\
c_db_nk   & $1.4$   & $-0.021$ & $1.75$ & $-0.025$    \\
c\_fb_nk  & $3.4$  & $ +0.20$  & $1.75$ & $-0.025$  \\
\midrule
ALMA average & $1.90$ & $-0.027$ & $1.90$ & $-0.027$ \\        
\bottomrule 
\toprule
\label{tab:IV}
\end{tabular}
\end{center}
\end{table}

\section{Discussion and conclusions}\label{sec:discussion}


We generate a family of semi-analytic models of an accretion flow with a bright orbiting spot to fit Sgr~A* millimeter data recorded by ALMA on April 11, 2017. Our models have a varying number of the degrees of freedom. The question which model represents the observational data best can be answered using either Bayesian or physical arguments.

\begin{figure*}[h!]
    \centering
    \includegraphics[width=0.76\linewidth,trim={0.cm 0.0cm 13.6cm 0.cm},clip]{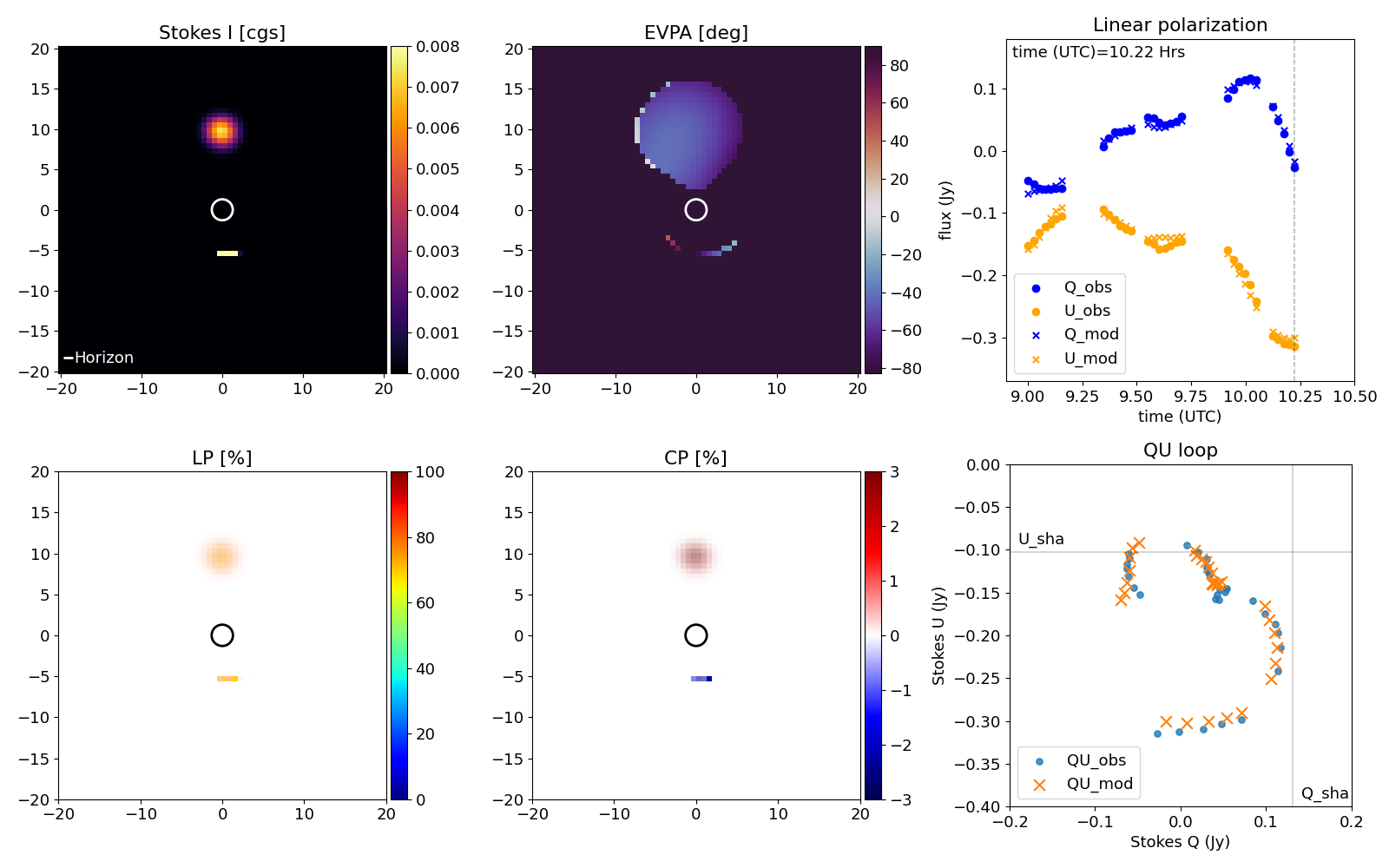} 
    \caption{Snapshot from the end of the light curve from our best-fitting model (nc_fb_nk), using $64^2$ resolution, for Stokes $\mathcal{I}$, EVPA, fractional linear polarization (LP), and fractional circular polarization (CP).}
    \label{fig:QU_loop_best}
\end{figure*}

The model that reproduces the data best, that is, corresponds to the lowest $\chi^2_{\rm eff}$ parameter, is shown in Fig.~\ref{fig:QU_loop_best}, made with a flipped $B_{\textrm{field}}$, no cooling, and a non-Keplerian orbit. Based on the $\chi^2_{\rm eff}$, the two best models are with a flipped $B_{\textrm{field}}$ (magnetic field vector arrow pointing toward the observer for viewing angles $i>90$\,deg). It is worth pointing out that the only difference between the two orientations comes from Faraday effects, and particularly, rotation. Additionally, in Table~\ref{tab:logz}, we report all the log-Evidence (marginal likelihood) from all models. Most models produce similar values, with two exceptions: nc\_fb\_nk, with a significantly higher value, and the two cooling and non-Keplerian models with a lower value. 

Based on physical arguments, the models that describe the observations best are those that in addition to Stokes $\mathcal{Q}$ and $\mathcal{U}$ also show some consistency with the observed Stokes $\mathcal{V}$ \citepalias[also see Appendix E of][]{W22}. The model that satisfies this requirement particularly well is still the one without cooling and no-Keplerian orbits. All models with a default magnetic field orientation fail to recover the Stokes $\mathcal{V}$ variability. We also find a mismatch between the modeled and observed Stokes $\mathcal{I}$ during the flare, but this is expected. Stokes $\mathcal{I}$ during the flare could be dominated by the background emission, which may be changing as well. The observed total intensity light curve appears to recover from the minimum following the X-ray flare, steadily increasing the flux density on a timescale of $\sim$2\,h, which may be interpreted as the accretion disk cooling down after the energy injection \citep{Wielgus2022_LC}. Alternatively, Stokes $\mathcal{I}$ during the flare may be also influenced by spot properties that we did not include in our simple model, such as the shearing of the spot or other spot interaction with the background flow \citep[e.g.,][]{Tiede2020}. 

In observational data, the $\mathcal{Q}-\mathcal{U}$ loop disappears on timescales of about $\sim2$\,h similar to the aforementioned total intensity recovery timescale. Thus, it might be argued that the best spot model synchrotron cooling timescale, $T_{\textrm{cool}}$, should also be close to 2\,h. Given our best-fit values for $\Theta_{0,s}$ and $B_{0,s}$ we calculated $T_{\textrm{cool}}$  using (\citetalias{W22} and \citealt{moscibrodzka:2011})
\begin{equation}
    T_{\textrm{cool}}= 11\,{\rm h} \left(\frac{B}{10G}\right)^{-2} \left(\frac{\Theta}{50}\right)^{-1} \,.
\label{eq:tcool}
\end{equation}
In Table~\ref{tab:logz}, we report the $T_{\textrm{cool}}$ values for all spot models. If the disappearance of the spot is due to radiative cooling, then there is only one model that matches this observational constraint reasonably well, the Keplerian model with cooling and a flipped magnetic field ($T_{\textrm{cool}}=$1\,h). However, we assumed that the electrons in the spot have a thermal distribution function. A different assumption, for instance, a nonthermal electron distribution function, could lead to different $T_{\textrm{cool}}$. 

Regardless of their details, all spot models show consistently similar best-fitting viewing angles $i \in(155,160)$ deg and a spot orbital radius $r_s\in(9,12.5)M$. The viewing angle close to 180\,deg determines a clockwise hot-spot rotation on the sky. This value is also fully consistent with the findings of \citetalias{W22} and other independent works \citep{gravity:2018,gravity:2020a,gravity:2020b,eht:2022_paperV,gravity2023}. 
For all models (with one exception), the preferred period is approximately 90 minutes ( 155 minutes for the exception), which is longer than the 70 minutes estimated by \citetalias{W22}. We fit to a data segment starting at an earlier time than \citetalias{W22}, and therefore, a possible interpretation of this discrepancy would be a decrease in the period with time that could result from a non-negligible radial velocity component of the hot spot toward the black hole, for example.

For the expected orbital radii, the period (Eq.~\ref{eq:P}) only weakly depends on the black hole spin. Although we obtain a black hole spin estimate of high values ($a_* \sim 0.8-0.9$) in all cases, we did not consider negative spins in our priors. We therefore missed some of the parameter space. A brief analysis of this estimation is available in Appendix \ref{app:spin}. A convincing independent estimate of the \sgra spin in the literature is lacking as well, so that we cannot compare our value to it. All models except for c\_fb\_nk also consistently show PA\,$\in[-10,10]$ deg (or $[170,190]$ deg because the electric vector position angle (EVPA) has an ambiguity of 180 deg ambiguity, meaning that the magnetic field orientation (and, if it is nonzero, also the black hole spin) may have a specific orientation on the sky. The PA corresponds to the position angle of the spin vector projected on the observer's screen, measured east-of-north (positive PA rotates the image counterclockwise). \citetalias{W22} reported a PA of 57\,deg, following a 32 deg counterclockwise rotation of the EVPA, which they attributed to the external Faraday screen. The existence of the external Faraday screen has recently been challenged by \citet{Wielgus2023}, and it is possible that the hot spot orbiting at $r_{s}\approx 11 M$ is located outside of the internal Faraday screen zone. In this case, no rotation should be applied, and the estimate of \citetalias{W22} becomes PA = 25\,deg. This value, given an uncertainty level of tens of degrees is close to our estimated range, which is acquired in a more systematic manner.

\begin{table}[tbh!]
\begin{center}
     \caption{Evidence ($\log Z \pm0.3$), $\chi_{\textrm{eff}}^2$, period, and cooling timescale $T_{\rm cool}$ (computed using Eq.~\ref{eq:tcool}) for all models.}
\begin{tabular}{lcccc}\toprule
 \midrule 
Model ID & Evidence & $\chi_{\textrm{eff}}^2$ &Period (min) & $T_{\textrm{cool}}$ (h) \\ 
\midrule    

nc\_db\_k  & $-63.6$    & $1.08$ &$87.8$& $70$   \\  
nc\_fb\_k & $-62.3$     & $0.54$  &$87.8$& $54$  \\
c\_db\_k   & $-63.1$    & $1.10$  &$87.2$&  $155$  \\
c\_fb\_k  & $-65.4$      & $1.22$ &$87.2$& $1$    \\
nc\_db\_nk  & $-67.7$   & $1.08$ &$91.8$& $93$  \\
nc\_fb\_nk & $-50.4$   & $0.34$  &$91.2$&  $77$  \\
c_b_nk   & $-73.6$   & $1.02$  &$92.3$& $52$   \\
c\_fb_nk  & $-92.8$   & $2.05$ &$155.5$& $52$    \\
\bottomrule 
\toprule
\label{tab:logz}
\end{tabular}
\end{center}
\end{table}


All models consistently show that our spot parameters should be in an approximate range of 
$n_{\rm 0,s}\in(10^6,3\times10^7)$ ${\rm particles/cm^3}$, $\Theta_{\rm 0,s} \in(15,70)$, $B_{\rm 0,s}\in(2,40)$\,G.
Additionally, the non-Keplerian models prefer $K_{\rm coef}\in(0.5,1)$, and in particular, $(0.65-0.85)$.
Subsequently, the properties of these hot spots might be found in more first-principle numerical (e.g., GRMHD or GRPIC) simulations of ADAFs. 
The GRMHD simulations of magnetically arrested disks \citep[MADs; ][]{Narayan2003} are known to show large-scale disk inhomogeneities threaded by vertical magnetic fields.
These structures, called flux tubes, are associated with magnetosphere reconnection and magnetic (flux) saturation of the black hole \citep{dexter20,porth21}. 
Our preliminary experiments with two-temperature 3D GRMHD simulations of MADs do show hot spots with properties matching our estimates of the spot parameters 
(Jim{\'e}nez-Rosales et al. in prep.). 
The flux tube scenario in MAD systems could be consistent with the $\mathcal{Q}-\mathcal{U}$ loops, as recently demonstrated by \cite{Najafi-Ziyazi}. Moreover, in the simulations, the flux eruption is usually associated with reconnection of the magnetosphere, which may explain the X-ray emission preceding the loop.
MAD disks have sub-Keplerian orbital profiles, in our terminology, averaged MADs $K_{\rm coef}\approx0.5-0.8$; see \citealt{Begelman:2022}, \citealt{Conroy:2023kec}, which agrees very well with the values we report ($0.65-0.85$). 
Even though flux tubes are a strong candidate for explaining the occurrence of observable hot spots (\citealt{Ripperda2022},\citetalias{W22}), we note that there are other possibilities, such as the formation of plasmoids \citep{ripperda20,elmellah23,vos23}.
Particularly when the orbital motion is super-Keplerian \citep[based on the infrared data analysis by, e.g.,][]{matsumoto20,Aimar2023}, this could be a more promising interpretation.

Fitting the background provided interesting results as well. First, the default $B_{\rm field}$ models produced more naturally correlated posteriors for the plasma parameters, considering the synchrotron radiation mechanism.  
Furthermore, the models with a magnetic field pointing toward the observer, which seem to produce better results for the spot fitting, have Stokes $\mathcal{V}_{\textrm{BG}}$ with the opposite sign to $\mathcal{V}_{\textrm{sha}}$ and the ALMA average value. The default $B_{\rm field}$ models, on the other hand, not only match the sign of Stokes $\mathcal{V}_{\textrm{BG}}$, but also the values for both Stokes $\mathcal{I}_{\textrm{BG}}$ and $\mathcal{V}_{\textrm{BG}}$. 
A flipped magnetic field for the disk, or even better, a separate morphology entirely for the spot and disk, might be the optimal solution.
Another possibility is that the shadow component has a negative circular polarization due to the gravitational lensing effect, which might be captured if our background model had a higher level of complexity (a similar change in the circular polarization sign of the direct and lensed image of the spot is visible in the bottom right panel in Fig.~\ref{fig:QU_loop_best}).
Overall, the background fitting can consistently reproduce the Stokes $\mathcal{Q}_{\textrm{sha}}$, $\mathcal{U}_{\textrm{sha}}$ dictated by the spot models, which is its main purpose. At a second level, the current setup is capable of producing appropriate values for the Stokes parameters, but it remains somewhat dissociated from the spot component. In a more dedicated study, a combined disk-spot model using our two components might match all Stokes parameters from the observations.

\begin{acknowledgements}
We thank Jongseo Kim and the anonymous EHT Collaboration internal reviewer for helpful comments. This publication is a part of the project Dutch Black Hole Consortium (with project number NWA 1292.19.202) of the research programme of the National Science Agenda which is financed by the Dutch Research Council (NWO). MM, AJR and AY acknowledge support from NWO, grant no. OCENW.KLEIN.113. MM also acknowledges support by the NWO Science Athena Award. MW acknowledges the support by the European Research Council advanced grant “M2FINDERS - Mapping Magnetic Fields with INterferometry Down to Event hoRizon Scales” (Grant No. 101018682). This paper makes use of the ALMA data set ADS/JAO.ALMA\#2016.1.01404.V; ALMA is a partnership of ESO (representing its member states), NSF (USA) and NINS (Japan), together with NRC (Canada), NSC and ASIAA (Taiwan), and KASI (Republic of Korea), in cooperation with the Republic of Chile. The Joint ALMA Observatory is operated by ESO, AUI/NRAO and NAOJ.
\end{acknowledgements}
\balance
\bibliographystyle{aa} 
\bibliography{library.bib} 

\newpage

\begin{appendix}
\section{Resolution}\label{app:res}

\begin{figure*}
    \centering
    \includegraphics[width=0.371\linewidth,trim={0.1cm 0.0cm 0 0.1cm},clip]{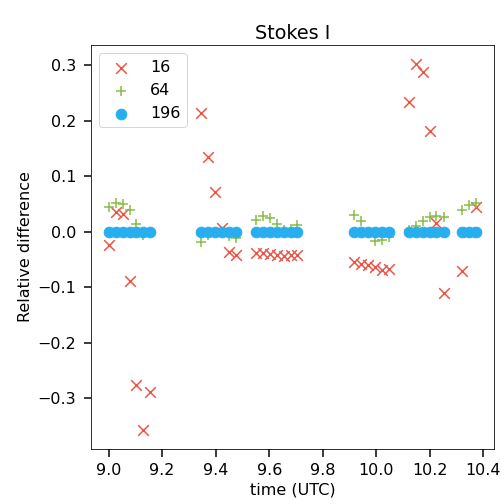}
    \includegraphics[width=0.371\linewidth,trim={0.1cm 0.0cm 0 0.1cm},clip]{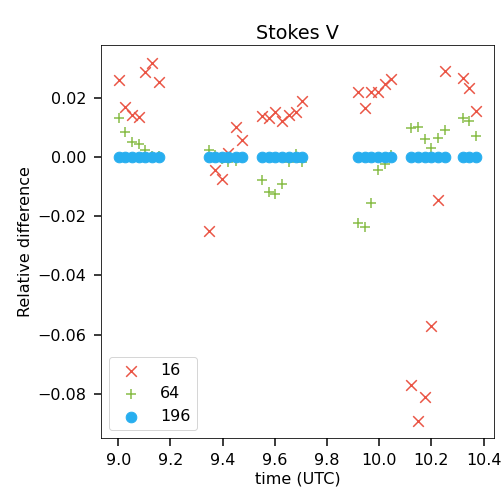} \\
    \includegraphics[width=0.371\linewidth,trim={0.1cm 0.0cm 0 0.1cm},clip]{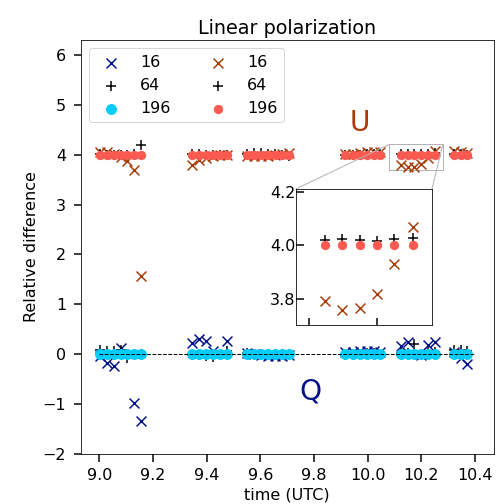}
    \includegraphics[width=0.371\linewidth,trim={0.1cm 0.0cm 0 0.1cm},clip]{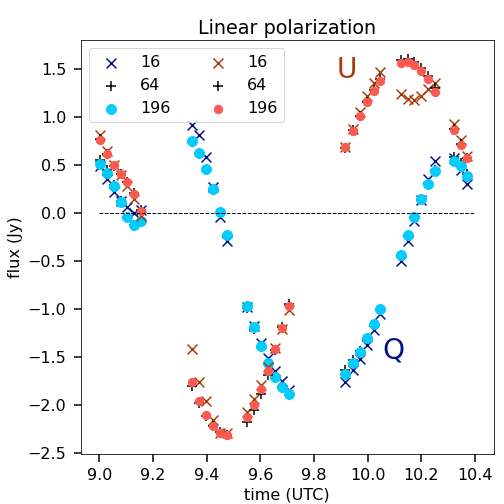}
    \caption{Relative differences between three different \texttt{ipole}  image resolutions ($16^2$, $64^2$, and $196^2$), for all Stokes parameters. In the case of Stokes $\mathcal{U}$, we have shifted the curve 4 units higher for readability reasons. The bottom right plot shows the full light curves for Stokes $\mathcal{Q}$ and $\mathcal{U}$, since fractional differences can be misleading for values close to zero.} 
    \label{fig:res}
\end{figure*}

To obtain the results presented in this paper on a reasonable timescale, we had to moderately reduce the image resolution in the ray-tracing part of \bipole from the more typical $400^2$ or $196^2$ to $64^2$. In what follows, we demonstrate that this does not affect the simulated light curves significantly, especially for Stokes $\mathcal{Q}, \mathcal{U}$. The latter is particularly important for our algorithm, which relies on fitting $\mathcal{Q}$ and $\mathcal{U}$ alone. 

In Fig.~\ref{fig:res} we show the relative difference (in percentile) of the light curves using three different resolutions ($16^2$, $64^2$, and $196^2$) for all Stokes parameters using a random combination of model parameters to create a test light curve that exhibits some structure (in essence, excluding models with constant light curves). Stokes $\mathcal{I}$ exhibits a maximum deviation of $\sim 5\%$ and Stokes $\mathcal{V}$ of $\sim 2\%$. For Stokes $\mathcal{Q}$ and $\mathcal{U}$ (see lower panels in Figure~\ref{fig:res}), we calculated the average deviation for $|\mathcal{Q}$, $\mathcal{U}| > 0.01$. This amounts to $2\%$ for $\mathcal{Q}$ and $1.5\%$ for $\mathcal{U}$. These values are lower than the assumed error budget (the thermal error is $e_t=0.01$, and the systematic error is $e_s=0.02$).

\begin{figure*}
    \centering
    \includegraphics[width=0.731\linewidth,trim={0.1cm 0.0cm 0 0.1cm},clip]{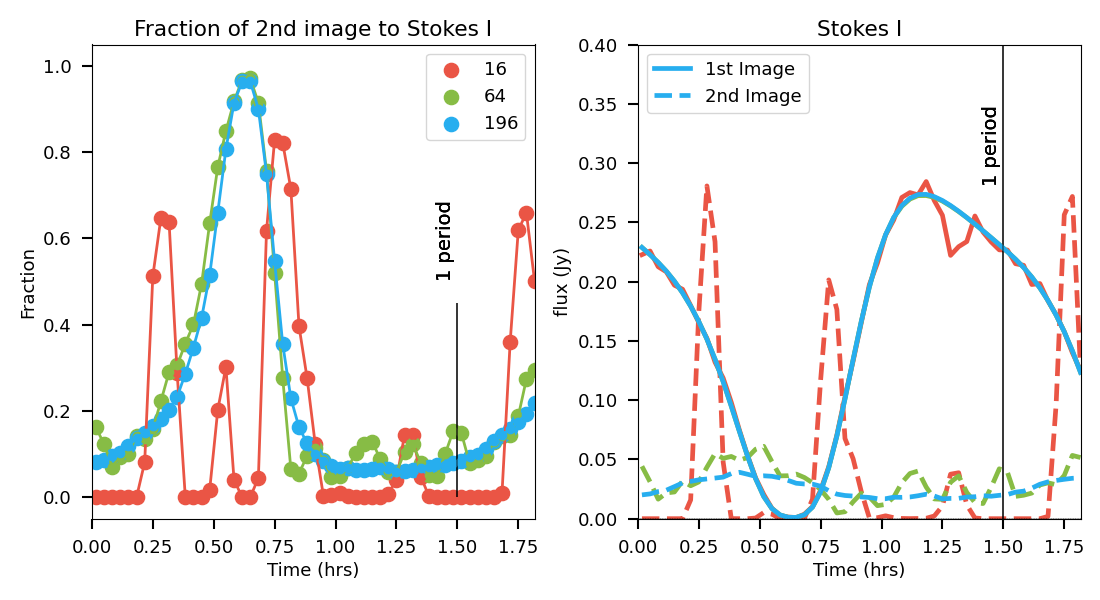}
    \caption{Comparison of emission from the first and the secondary image for different image resolutions. Left: Fraction of intensity contributed by the secondary image to the total Stokes I in a full period. The colors denote different image resolutions ($16^2$, $64^2$, and $196^2$). Right: Absolute flux of the the primary (solid line) and secondary (dashed line) image for the same three resolutions.}
    \label{fig:res_Ifrac}
\end{figure*}

The second test we performed determined the relation of the first to the secondary image, and how in particular the latter is affected by the resolution. The results are shown in Fig. \ref{fig:res_Ifrac}, where it is clear that with a resolution of $16 \times 16$, it is really hard to capture the contribution of the secondary image to Stokes $\mathcal{I}$ precisely, but at $64 \times 64$, the light curve resembles the higher resolution.

\section{Spin estimates}\label{app:spin}

Although the spin estimates acquired in this work are simply a byproduct of our fitting algorithm and not the main focus of our analysis, we discuss them in some more detail.

\begin{figure*}
\includegraphics[width=0.321\linewidth,trim={0.1cm 0.0cm 0 0.1cm},clip]{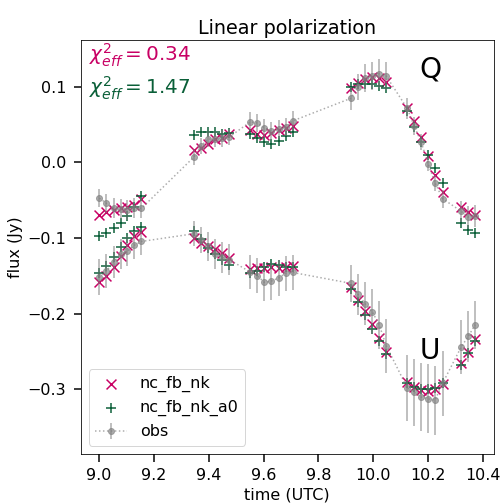}
\includegraphics[width=0.321\linewidth,trim={0.1cm 0.0cm 0 0.1cm},clip]{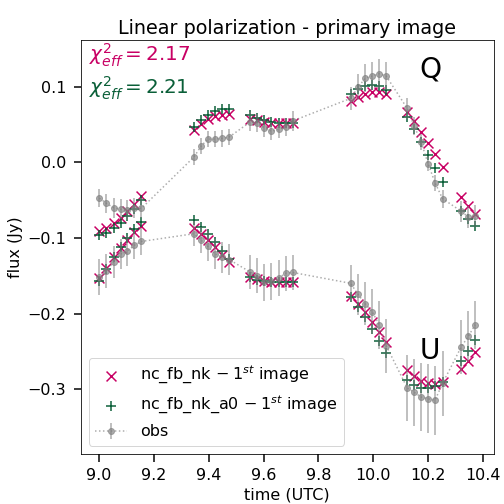}\includegraphics[width=0.321\linewidth,trim={0.1cm 0.0cm 0 0.1cm},clip]{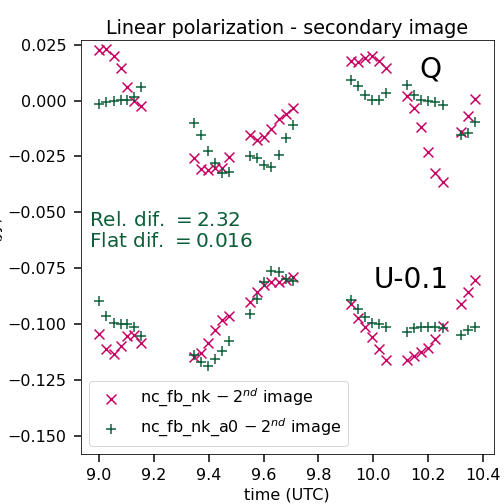}
    \caption{Difference in $\mathcal{Q,U}$ light curves from the best-bet model (nc\_fb\_nk), and the same model, but with the spin fixed to $a_*=0$ (nc\_fb\_nk\_a0). Left: Full light curve. Middle: Primary image. Right: Secondary image.}
    \label{fig:spin_simple}
\end{figure*}

In the first panel of Fig.~\ref{fig:spin_simple}, we present what happens to our best-bet model if we keep all parameters the same and only change the spin to $a_*=0$. While the difference is not dramatic (the model still produces $\chi^2_{\rm eff}\sim1.5$), it is certainly noticeable. To answer the question of what might cause this, we refer to panels 2 and 3, where we plot the light curves for the primary and secondary image separately. The middle panel clearly shows that the secondary image is necessary to produce the extreme similarity of our best-bet model with the data because the fit quality drops to $\chi^2_{\rm eff}>2$. Another striking observation is that the difference between the models now is negligible. This most likely comes from the slight change in the period (affected by spin for a fixed orbital radius and Keplerianity parameter) from $92.5$ to $89.5$\,min. This is further supported in the right panel of Fig.~\ref{fig:spin_simple}, where the two light curves, corresponding to the secondary image alone, are shown. The differences are significant, more than $200\%$ in relative terms and over $0.015$\,Jy in terms of the absolute flux density. This explains the spin preference, even though the spot is located so far way from the black hole. The dominant effect comes from the secondary image, which is formed by photons approaching the photon shell, which is located much closer to the black hole. A further analysis of the secondary image behavior and its dependence on the black hole spin necessitates a separate future work.   
\end{appendix}

%
%

\end{document}